\documentclass[12pt]{article}
\usepackage{bbm}
\usepackage{amssymb}
\newcommand{\nc}[2]{\newcommand{#1}{#2}}
\renewcommand{\theequation}{\arabic{section}.\arabic{equation}}
\nc{\Z}{\mathbbm{Z}}
\nc{\N}{\mathbbm{N}}
\nc{\R}{\mathbbm{R}}
\nc{\T}{\mathbbm{T}}
\nc{\C}{\mathbbm{C}}
\nc{\cA}{{\cal A}}
\nc{\cO}{{\cal O}}
\nc{\cH}{{\cal H}}
\nc{\id}{{\bf 1}}
\nc{\beq}{\begin{equation}}
\nc{\eeq}{\end{equation}}
\nc{\bea}{\begin{eqnarray}}
\nc{\eea}{\end{eqnarray}}
\nc{\beas}{\begin{eqnarray*}}
\nc{\eeas}{\end{eqnarray*}}
\nc{\ba}{\begin{array}}
\nc{\ea}{\end{array}}
\nc{\lra}{\longrightarrow}
\nc{\ra}{\rightarrow}
\nc{\ot}{\otimes}
\nc{\ci}{\circ}
\nc{\bpf}{{\em Proof:\enspace\enspace}}
\newtheorem{pr}{Proposition}
\newtheorem{de}{Definition}
\newtheorem{them}{Theorem}
\newtheorem{lem}{Lemma}
\newtheorem{conj}{Conjecture}

\newtheorem{re}{Remark}

\nc{\bpr}{\begin{pr}} 
\nc{\bth}{\begin{them}}
\nc{\ble}{\begin{lem}} 
\nc{\bco}{\begin{corollary}} 
\nc{\bre}{\begin{re}} 
\nc{\bex}{\begin{example}} 
\nc{\bde}{\begin{de}} 
\nc{\ede}{\end{de}} 
\nc{\esa}{\end{satz}} 
\nc{\epr}{\end{pr}} 
\nc{\ethe}{\end{them}} 
\nc{\ele}{\end{lem}} 
\nc{\eco}{\end{corollary}} 
\nc{\ere}{\end{re}} 
\nc{\eex}{\end{example}}
\nc{\epf}{\hfill{$\square$}}

\begin{document}

\title{\bf Spectral triples and differential calculi related to the Kronecker
foliation} \author{R.~Matthes\\
\em Fachbereich Physik, Technische Universit{\"a}t Clausthal,\\
\em Leibnizstra{\ss}e 10, 38678 Clausthal-Zellerfeld\\[2ex]
O.~Richter and G.~Rudolph\\
\em Institut f{\"u}r Theoretische Physik, Universit{\"a}t Leipzig,\\
\em Augustusplatz 10/11, 04109 Leipzig, Germany}

\thispagestyle{empty}

\date{}

\maketitle

\abstract{Following ideas of Connes and Moscovici, we describe two spectral
triples related to the Kronecker foliation, whose generalized Dirac operators
are related to first and second order signature operators. 
We also consider the corresponding differential calculi $\Omega_D$, which
are drastically different in the two cases. As a side-remark, we give a
description of a known calculus on the two-dimensional noncommutative torus
in terms of generators and relations.}\\[2ex]

\section{Introduction}

In Connes' approach to noncommutative differential geometry, the notion
of a spectral triple plays an essential role, see~\cite{co8.5}. It encodes 
the differential
and Riemannian structure of the noncommutative space as well as its dimension.
From the physical point of view, spectral triples have been used to
construct unified field theoretical models, in particular the standard
model (see~\cite{co8.5},~\cite{colo}), and also models including gravitation 
(\cite{chco}, \cite{kawa-1}, \cite{ka6}). 
From the mathematical point of view, only a few types of noncommutative
spaces have been used in these examples: commutative 
algebras of smooth functions on a manifold~\cite{co8.5}, finite dimensional 
algebras~(for a classification
of spectral triples in this case see \cite{kraj} and \cite{ps98}) and products
of both. In \cite{scgm9} it was shown that it is not straightforward to define 
spectral triples related to covariant
differential calculi on quantum groups. Explicit examples of spectral triples
have also been described for the irrational rotation algebra and higher
dimensional noncommutative tori ~\cite{co8.5}, \cite{varil1}. For these
examples,  the data of the triple were chosen according to physical needs or
taking advantage of special structures available in the underlying algebra. 
An important part of the information needed for physical purposes is the
explicit form of the differential calculus of a spectral triple. Such calculi
have been analyzed in the above-mentioned cases (\cite{co8.5},~\cite{colo},
\cite{kppw}, \cite{mrw}). 
In
\cite{mrw1} it has been shown that the extra structure of a finitely
generated projective module allows to introduce the graded algebra
of differential-form-valued endomorphisms which gives a natural mathematical
language to build unified field theoretical models in the spirit of
the Mainz-Marseille approach \cite{cefs1}.
In \cite{fgr1} and \cite{fgr2}, the
notion of spectral triple itself has been modified and enriched using ideas
from supersymmetric quantum theory.  One arrives at noncommutative structures
generalizing classical geometrical structures (Riemannian, symplectic,
Hermitian, K{\"a}hler, \dots structures). Physical hopes are mainly directed to
superconformal field theories (with noncommutative target spaces). 

Recently, see~\cite{como.5}, Connes and Moscovici have described a method which
makes it possible to construct spectral triples in a systematic way
for crossed product algebras related to foliations. Let $(M,{\cal F})$ be a
regular foliation of a smooth manifold $M$ with Euclidean structures on both
the corresponding distribution and the normal bundle. There is an associated 
spectral triple for the crossed product algebra $C^\infty(M)\rtimes\Gamma$,
where $\Gamma$ is a group of diffeomorphisms preserving these
structures. The corresponding Dirac operator is a hypoelliptic operator
which is closely related to the signature operator of the foliated manifold.
This signature operator is a modification of the standard signature
operator in differential geometry, see~\cite{gilkey1}. 

In this paper, we construct explicitly two spectral triples related to the 
Kronecker
foliation. We choose as diffeomorphism group the group $\R$ which defines
the foliation by its action on $\T^2$ and obviously preserves
natural translation invariant Euclidean structures. Thus we 
arrive at the algebra  $C^\infty(\T^2)\rtimes \R$, whose $C^*$ version is known
to be Morita equivalent to the irrational rotation algebra
(noncommutative torus), see \cite{rieffel1}, \cite{varil1}.
The Dirac operator of the
first spectral triple (which has dimension 2) is closely related to the
ordinary signature operator on $\T^2$. For the construction of the second
triple (of dimension three) we follow the strategy proposed 
in~\cite{como.5}.
The corresponding signature operators
and henceforth also the Dirac operators can be diagonalized explicitly in
both cases. 
Then we pass to the differential calculi associated to the
spectral triples constructed before. It turns out that for the triple related 
to the first order signature
operator the differential calculus can be completely determined. Restricted to
$C^\infty(\T^2)$ it projects down to the de Rham calculus on $\T^2$. The
analysis of the differential calculus for the second triple turns out to be
much more involved. We show that for the restriction of this triple to
the subalgebra $C^\infty(\T^2)$ (i.e. choosing the trivial diffeomorphism
group) the corresponding one forms give just 
the universal calculus on $C^\infty(\T^2)$.

In the appendix we have added the explicit description of the differential 
calculus
for the spectral triple related to the irrational rotation algebra,
see~\cite{co8.5,varil1}, which has properties similar to the calculus associated
to the linear signature operator.

\section{The spectral triple related to a foliation}
\setcounter{equation}{0}

For the convenience of the reader, we recall here the definition of
a spectral triple and the differential calculus
related to such a triple (\cite{co8.5},\cite{varil1}): 
\bde
A spectral triple $(A,\cH,D)$ consists of a $*$-algebra $A$,
a Hilbert space $\cH$ and an unbounded operator $D$ on $\cH$,
such that
\begin{enumerate}
\item[(i)] $A$ acts by a $*$-representation $\pi$ in the algebra $B(\cH)$ 
of bounded operators on $\cH$,
\item[(ii)] the commutators $[D,\pi(a)],~a\in A$, are bounded and
\item[(iii)] the operator $D$ has discrete spectrum with finite
multiplicity.
\end{enumerate}
A spectral triple is said to have dimension $n$, if the 
eigenvalues (with multiplicity) $\mu_k$ of $|D|$ fulfil
$\lim_{k\to\infty}\frac{\mu_k}{k^{1/n}}=C\neq 0$.  
\ede
We will have no need to refer to gradings or real structures usually
included in the definition of a spectral triple, and also not to more
general notions of dimension. 

The representation $\pi$ of $A$ in $B(\cH)$ can be extended to a representation
$\pi^*:\Omega(A)\lra B(\cH)$ of the universal differential calculus $\Omega(A)$
by 
\[
\pi^n(\sum_ka_0^kda_1^k\cdots da_n^k)=\sum_k\pi(a_0^k)[D,\pi(a^1_k)]\cdots 
[D,\pi(a^k_n)].
\]
If $J_0:=\oplus_n\ker \pi^n$, then $J:=J_0+dJ_0$ is a differential ideal, and
one arrives at the differential calculus $\Omega_D(A)$, 
\[
\Omega_D^n(A):=\Omega^n(A)/J.
\]
Note that, if $\pi$ is faithful, there are isomorphisms
\beq\label{o1}
\Omega^1_D(A)\simeq \pi^1(\Omega^1(A))
\eeq
and
\beq\label{o2}
\Omega^2_D(A)\simeq \pi^2(\Omega^2(A))/\pi^2(dJ_0^1).
\eeq

Now we review shortly the procedure given in~\cite{como.5}, which relates a 
spectral triple to a regular foliation of a smooth manifold. Let $M$ be a
compact manifold with a foliation given by an integrable distribution 
$V\subset TM$. The normal bundle of the foliation is
$N:=TM/V$, with canonical projection $\rho:TM\to N$.
Assume further that both $V$ and $N$ are equipped with
Euclidean fibre metrics and with an orientation (i.e. there are
distinguished nowhere vanishing sections $\omega_V$,
$\omega_N$ of the exterior bundles $\bigwedge^vV$, 
$\bigwedge^nN$ ($v=\dim V,~n=\dim N$)).
Furthermore, $\omega_V$ and $\omega_N$ also define a nonvanishing section
of $\bigwedge^v V^*\ot\bigwedge^nN^*\simeq \bigwedge^{v+n}T^*M$,
i.e. a volume form on $M$. The bundle of interest for us is
\[
E=\bigwedge V_\C^*\otimes \bigwedge N_\C^*.
\]
Obviously, the metrics on $V$ and $N$ give rise to Hermitian metrics on
$\bigwedge V_\C^*$ and $\bigwedge N_\C^*$ and thus also on $E$.
The orientations $\omega_V$ and $\omega_N$ can be mapped by means of the
metrics to sections $\gamma_V$ of $\bigwedge^vV^*_\C$ and $\gamma_N$
of $\bigwedge^nN^*_\C$
which can be used, together with the metrics, to define
an analogue of the Hodge star on the exterior bundles $\bigwedge
V_\C^*$ and $\bigwedge N_\C^*$. We choose a variant of the $*$-operation
such that $*_{V_\C}^2=1$ and $*_{N_\C}^2=1$, i.e. $*_{V_\C}$ and $*_{N_\C}$
can be considered as $\Z_2$-grading operators (cf. \cite{geve}).\\
Thus, the space of sections of $E$ has a natural
inner product, and we denote by ${\cal H}=L^2(M,E)$ the Hilbert space of square
integrable sections of this bundle. From now on, we always consider
complexified vector bundles, but omit the subscript $\C$.

In order to construct a generalized Dirac operator, a longitudinal
differential $d_L$ and a transversal differential operator $d_H$ have to be
defined. The differential $d_L$ is defined canonically by means of the Bott 
connection (\cite{bo}) given as the partial covariant derivative 
$\nabla:\Gamma(V)\times \Gamma(N)\lra \Gamma(N)$ defined by
\[
\nabla_XY=\rho\left([X,\tilde{Y}]\right)\enspace,
\]
for $X\in \Gamma(V),Y\in \Gamma(N)$ and $\tilde{Y}\in \Gamma(TM)$ 
such that $\rho(\tilde{Y})=Y$. 
By a standard procedure (using the
Leibniz rule and duality) $\nabla$ is extended to a differential
$d_L:\Gamma(E)\lra \Gamma(E)$ defined by linear mappings
$\Gamma(\bigwedge^kV^*\ot\bigwedge^lN^*)\lra
\Gamma(\bigwedge^{k+1}V^*\ot\bigwedge^lN^*)$,
\begin{eqnarray*}
d_L\alpha(X_0,\ldots,X_k)&=&
\sum_{i=0,\ldots,k}(-1)^i\nabla_{X_i}(\alpha(X_0,\ldots,\hat{X_i},\ldots,X_k))\nonumber\\
&+&
\sum_{i<j}(-1)^{i+j}\alpha([X_i,X_j],X_0,\ldots,\hat{X_i},\ldots,\hat{X_j},
\ldots,X_k),
\end{eqnarray*}
$X_i\in \Gamma(V)$.
Since the Bott connection is flat, we have $d_L^2=0$. 

In order to define a transversal differential operator one has to choose
a subbundle $H\subset TM$ complementary to $V$. This defines 
a bundle isomorphism $j_H:\bigwedge V^*\ot \bigwedge N^*\lra \bigwedge T^*M$ 
in the
following way: Let us denote by pr$_V^*$ and pr$_H^*$ the projections corresponding
to the decomposition $TM^*=V^*\oplus H^*$, by $\rho_H:H\to N$ the restriction
of $\rho$ to $H$ and by $\rho_H^*$ its transposed map.
Then $j_H$ is defined as the following composition:
\[
\bigwedge V^*\otimes\bigwedge N^*
\stackrel{{\rm id}\otimes\wedge\rho_H^*}{\longrightarrow}\bigwedge
V^*\otimes\bigwedge H^*
\stackrel{\wedge{\rm pr}_V^*\otimes\wedge{\rm pr}_H^*}{\longrightarrow}
\bigwedge T^*M\otimes\bigwedge T^*M
\stackrel{\otimes\to\wedge}{\longrightarrow} T^*M\enspace,
\]
where $\otimes\to\wedge$ denotes the replacement of the tensor product by the
wedge product. Now, the transversal operator $d_H$ is obtained from the
exterior differential $d$ by transporting with $j_H$ and
projecting to a certain homogeneous component:
$\bigwedge V^*\ot\bigwedge N^*$ has an obvious bigrading, and denoting by 
$\pi^{(r,s)}$ the projector to the homogeneous component of bidegree
$(r,s)$, one defines
\[
d_H\alpha=\pi^{(r,s+1)}(j_H^{-1}\ci d\ci j_H(\alpha))
\]
for $\alpha\in\Gamma(\bigwedge^rV^*\ot\bigwedge^sN^*)$. The operator $d_H$ is a graded
derivation of the $\Z_2$-graded algebra $\Gamma(\bigwedge V^*\ot\bigwedge
N^*)$.

In a foliation chart, $d_L$ and $d_H$ look as follows.
Let $(x^i,y^k)$, $i=1,\ldots,v$, $k=1,\ldots,n$ be local
coordinates of $M$ such that $x^i$ are
coordinates on the leaf (foliation chart).
The corresponding coordinate vector fields $(\frac{\partial}{\partial x^i},
\frac{\partial}{\partial y^k})$
form a local frame of $TM$ and $(\frac{\partial}{\partial x^i})$ 
a frame of $V$. The corresponding dual frame of $T^*M$ consists of the
differentials $(dx^i,dy^k)$. We define $\theta^i\in\Gamma(V^*)$
by $\theta^i(\frac{\partial}{\partial x^j})=\delta^i_j$
($i,j=1,\ldots,v$). It is immediate from the
definition of $N$ that the elements $n_k:=\frac{\partial}{\partial y^k}+V$
($k=1,\ldots,n$) form a local frame of $N$. The elements of the corresponding dual
frame of $N^*$ are denoted by $n^k$. Finally, we choose a local
frame $h_k$ of the transversal space $H$. This frame is fixed by 
assuming 
$\rho_H(h_k)=n_k$. 
This leads to
\[
h_k=h^i_k\frac{\partial}{\partial x^i}+\frac{\partial}{\partial
y^k}\enspace,
\]
with coefficient functions $h^i_k$ characterizing $H$.
Then, the elements $\theta^{i_1}\wedge\cdots\wedge\theta^{i_r}\ot
n^{j_1}\wedge\cdots\wedge n^{j_s}$ form a local frame of $E$, and one can
show that $d_L$ and $d_H$ are given by the following local formulae:
\[
d_L(\alpha_{i_1\ldots i_rj_1\ldots j_s}
\theta^{i_1}\wedge\cdots\wedge\theta^{i_r}\ot
n^{j_1}\wedge\cdots\wedge n^{j_s})=
\frac{\partial\alpha_{i_1\ldots i_rj_1\ldots j_s}}{\partial x^i}
\theta^i\wedge\theta^{i_1}\wedge\cdots\theta^{i_r}\ot n^{j_1}\wedge\cdots
\wedge n^{j_s}
\]
\bea\nonumber
&&d_H(\alpha_{i_1\ldots i_rj_1\ldots j_s}
\theta^{i_1}\wedge\cdots\wedge\theta^{i_r}\ot
n^{j_1}\wedge\cdots\wedge n^{j_s})=\\\label{dhloc}
&&(-1)^r\left(\frac{\partial }{\partial y^k}+
h^i_k\frac{\partial }{\partial x^i}\right)
\alpha_{i_1\ldots i_rj_1\ldots j_s}
\theta^{i_1}\wedge\cdots\wedge\theta^{i_r}
\ot
n^k\wedge n^{j_1}\wedge\cdots\wedge n^{j_s}
+\nonumber\\
&&\alpha_{i_1\ldots i_rj_1\ldots j_s}\sum_{t=1}^r\frac{\partial
h^{i_t}_k}{\partial x^l}
\theta^{i_1}\wedge\cdots\wedge\theta^l\wedge\cdots\wedge\theta^{i_r}\ot
n^k\wedge n^{j_1}\wedge\cdots\wedge n^{j_s}\enspace,
\eea
(where $\theta^l$ at position $t$ replaces $\theta^{i_t}$). The longitudinal
differential $d_L$ acts as a differential in leaf direction, whereas $d_H$
is a sum of a principal part, which differentiates in transversal
direction, and a zero order part.
As examples, let us give formulae for $d_H$ acting on functions,
(1,0)-, (0,1)- and (1,1)-forms:
\begin{eqnarray*}
d_Hf&=&h_k(f) n^k\enspace,\\
d_H(\alpha_i\theta^i)&=&-h_k(\alpha_i)\theta^i\ot n^k -
\alpha_i\frac{\partial h^i_k}{\partial x^j}\theta^j\ot n^k\enspace,\\
d_H(\alpha_kn^k)&=&h_l(\alpha_k)n^l\wedge n^k\enspace,\\
d_H(\alpha_{ik}\theta^i\wedge n^k)&=&-h_l(\alpha_{ik})\theta^i\ot n^l\wedge n^k
-\alpha_{ik}\frac{\partial h^i_l}{\partial x^j}\theta^j\ot 
n^l\wedge n^k\enspace,
\end{eqnarray*} 
($d_H(n^k)=0$). For the adjoint operators $d_L^*$ and $d_H^*$ (in ${\cal H}$) 
it is difficult to write down explicit formulae. One can show
\[
d_L^*\alpha=-*_Vd_L*_V +\enspace\mbox{term of order zero},
\]
where $*_V$ is the (partial) Hodge operator related to the Euclidean metric
and the orientation of $V$. Since $d_H^*$ lowers the $N^*$-degree one has for 
$\alpha\in \Gamma(\bigwedge^rV^*)\equiv\Gamma(\bigwedge^rV^*\ot \bigwedge^0N^*)$
\[
d_H^*\alpha=0\enspace.
\]
Explicit formulae for $d_H^*$ become rather complicated as, e.g., the 
case of (0,1)-forms shows:
\beq\label{dh*loc}
d_H^*(\alpha_i\theta^i)=-g_N^{kl}(h_k(\alpha_l)-\alpha_m{\Gamma_N}^m_{kl}
+\alpha_l(\frac{\partial h^i_k}{\partial
x^i}+\frac{1}{2}g_V^{ij}h_k({g_V}_{ij})))\enspace,
\eeq
where $g_N^{kl}=g_N(n^k,n^l)$, ${g_V}_{ij}=g_V(\frac{\partial }{\partial
x^i},\frac{\partial }{\partial
x^j})$, $g_V^{ij}=g_V(\theta^i,\theta^j)$ are the local components of the
fibre metrics (and their duals), and ${\Gamma_N}^m_{kl}$ are the
Christoffel symbols corresponding to ${g_N}_{kl}$.

In~\cite{como.5}, using $d_H$ and $d_L$, two differential operators were
introduced by 
\begin{eqnarray*}
Q_L&=&d_Ld_L^*-d_L^*d_L\\
Q_H&=&d_H+d_H^*\enspace,
\end{eqnarray*}
and the mixed signature operator $Q$ for $M$, acting on a form with
$N$-degree $\partial_N$, was defined by
\beq\label{qmix}
Q=Q_L(-1)^{\partial_N}+Q_H\enspace.
\eeq
As noted in \cite{como.5}, $Q$ is selfadjoint. Finally, a generalized 
Dirac operator $D$ is defined as the unique selfadjoint operator such 
that
\beq\label{d|d|}
D|D|=Q\enspace.
\eeq
If zero is not an element of the spectrum of
$Q$, it is given as
\beq\label{Q}
D=Q|Q|^{-1/2}=Q(Q^2)^{-1/4},
\eeq
as shows a straightforward argument using the spectral decomposition of $Q$.

One motivation for choosing a second order longitudinal part is the
following: the index of the signature operator should not depend on the choice 
of the transversal subbundle $H$. Usually, the index of a pseudodifferential 
operator only
depends on its principal symbol. However, 
as follows from the local formulae (\ref{dhloc}) and (\ref{dh*loc}), 
its principal part explicitly depends
on $H$, the dependence being in the coefficients of the partial derivatives
with respect to leaf coordinates. It turns out that one can get rid of this
dependence by introducing a modified notion of pseudodifferential operators
($\psi DO'$) which assigns a degree 2 to transversal coordinates and a degree
1 to longitudinal ones. To have a contribution also from $Q_L$, one has to
pass to a second order operator. In~\cite{como.5} a homotopy argument was given
to show that this does not affect the longitudinal signature class.

Let $\Gamma$ be any group of diffeomorphisms of $M$ which preserves the
distribution $V$ and the Euclidean metrics on both $V$ and $N$. Then 
$\psi\in\Gamma$ acts via the pull back as unitary operator 
$U^*_\psi$ on $\cH$, whereas
functions from $C^\infty(M)$ act there as multiplication operators.
The crossed product 
algebra $\cA:=C^\infty(M)\rtimes\Gamma$ can be defined as the $*$-subalgebra
of $B(\cH)$ generated by these two types of operators. Due to $U^*_\psi f=
(f\ci\psi)U^*_\psi$ every element of $\cA$
is a finite sum of elements $fU^*_\psi$.
Then we have, see~\cite{como.5}:
\bth\label{thm1}
$(\cA,{\cal H},D)$ is a spectral triple of dimension $v+2n$.
\ethe
Our main aim is to describe explicitly this spectral triple for the Kronecker
foliation of the two-torus.

\section{Spectral triples for the Kronecker foliation}
\setcounter{equation}{0}
\subsection{The crossed product algebra for the Kronecker foliation}

Let us start with some conventions and notations.
We consider the two-torus as the quotient $\T^2=\R^2/2\pi\Z^2$. Thus, we have
natural local coordinates $0<\vartheta_1,\vartheta_2<2\pi$. Consider the
$\R$-manifold $(\T^2,\R,\psi)$, with group action
\[
\psi: \T^2\times\R\to\T^2\enspace,
\]
given by
\[
\psi((\vartheta_1,\vartheta_2),t)=(\vartheta_1+at,\vartheta_2+b t)\enspace,
\]
with $a,b\in\R$ such that $a>0$, $a^2+b^2=1$ and $\theta=\frac{b}{a}$ being 
irrational. The foliation of $\T^2$ by the orbits of $\psi$ is called the
Kronecker foliation. It is well known, see~\cite{molino1}, that each leaf of
this foliation is diffeomorphic to $\R$ and lies dense in $\T^2$. 

The coordinate transformation
\begin{eqnarray*}
x&=&a\vartheta_1+b\vartheta_2\\
y&=&b\vartheta_1-a\vartheta_2
\end{eqnarray*}
is orthogonal and leads to coordinates $(x,y)$ of a foliation chart.
In these coordinates, $\R$ acts as follows
\[
\psi((x,y),t)=(x+t,y)\enspace.
\]
To be more precise, this is the lifted action of $\R$ on $\R^2$,
applied to global coordinates $(x,y)$ obtained from global coordinates
$(\vartheta_1,\vartheta_2)$ by the orthogonal transformation. 

It is well known, see~\cite{renault1}, that associated to the action of a locally
compact group $K$ on a manifold $M$ there is a transformation groupoid $G$. 
For the Kronecker foliation, we have $G=\T^2\times\R$ with range and
source maps $r$ and $s$ given by
\begin{eqnarray*}
r(p,t)&=&\psi(p,t)\\
s(p,t)&=&p\enspace,
\end{eqnarray*}
$p\in\T^2$, the space of units being $\T^2$. The associated crossed product
algebra  
\[
{\cO}=\cO(\T^2)\rtimes\R\enspace
\]
is the $*$-algebra generated 
by the unitary operators $U_1$, $U_2$ and $V_t$ acting in the Hilbert space
$L^2(\T^2)$ given by
\begin{eqnarray}\label{gen} (U_1\xi)(\vartheta_1,\vartheta_2)&=&{\rm e}^{{\rm
i}\vartheta_1}\cdot\xi(\vartheta_1,\vartheta_2)\nonumber\\
(U_2\xi)(\vartheta_1,\vartheta_2)&=&{\rm e}^{{\rm
i}\vartheta_2}\cdot\xi(\vartheta_1,\vartheta_2)\nonumber\\
(V_t\xi)(\vartheta_1,\vartheta_2)&=&\xi(\vartheta_1+at,\vartheta_2+b t)\enspace,
\end{eqnarray}
$\forall\xi\in L^2(T^2)$. Let $e_{kl}={\rm e}^{{\rm
i}(k\vartheta_1+l\vartheta_2)}$ ($k,l\in\Z$) be the basis of trigonometric
polynomials of $L^2(\T^2)$. Obviously from (\ref{gen}) it follows that
\bea
U_1e_{kl}&=&e_{k+1,l}\nonumber\\
U_2e_{kl}&=&e_{k,l+1}\nonumber\\
V_te_{kl}&=&{\rm e}^{{\rm i}(ak+bl)t}e_{kl}\enspace.\label{gena}
\eea
It is now immediate to show
\bpr
The unitary operators $U_1$, $U_2$, $V_t$ satisfy
\bea\label{u1u2}
U_1U_2&=&U_2U_1,\\
V_tU_1&=&{\rm e}^{{\rm i}at}U_1V_t,\label{vu1}\\
V_tU_2&=&{\rm e}^{{\rm i}bt}U_2V_t,\label{vu2}\\
V_tV_s&=&V_{t+s},~t,s\in\R.\label{vv}
\eea
\epr
\bre
For rational $\frac{a}{b}=\frac{m}{n}$, $m,n$ relative prime, there is an
additional relation
\[
V_{2\pi\sqrt{m^2+n^2}}=V_0=1\enspace;
\]
$2\pi\sqrt{m^2+n^2}$ is the smallest value of $t$ such that $V_t=V_0=1$ and any
other such $t$ is an integer multiple of it. 
\ere   
\bpr\label{alg}
The $*$-algebra $\cO(\T^2)\rtimes\R$ is isomorphic to 
$\C\langle u_1,u_2,v_t\rangle/J$,
where $\C\langle u_1,u_2,v_t\rangle$ is the free associative unital
$*$-algebra generated by $u_1,u_2$ and $v_t,~t\in\R$, and $J$ is the 
$*$-ideal generated by (\ref{u1u2})--(\ref{vv}) and unitarity
conditions for the generators.
\epr
\bpf
By universality of $\C\langle u_1,u_2,v_t\rangle/J$,
there exists a homomorphism $\pi$ of this algebra onto $\cO(\T^2)\rtimes\R$ 
sending the corresponding generators into each other. Now, by 
(\ref{u1u2})--(\ref{vv}), a general element $a$ of $\C\langle
u_1,u_2,v_t\rangle/J$
is a linear combination of the monomials $v_{t_j}u_1^ku_2^l$, for some
$t_j\in\R$ (with $j\neq j'~\Rightarrow t_j\neq t_{j'}$), $k,l\in \Z$, 
$a=\sum a_{jkl}v_{t_j}u_1^ku_2^l$ (finite sum). 
We first show
that all $V_t$ are independent. $V_t=\sum_jb_jV_{t_j}$ is equivalent to 
$1=\sum b_je^{i(ka+bl)(t_j-t)}$, $\forall k,l$. Since $\frac{a}{b}$ is
irrational, $\{ak+bl|k,l\in\Z\}$ is dense in $\R$,
and one can conclude $1=\sum b_je^{ix(t_j-t)}$ $\forall x\in\R$.
But 1 and ${\rm e}^{{\rm i}x(t_j-t)}$ are orthogonal as almost-periodic 
functions, see~\cite{Akh/Gl}, and therefore linearly independent.
Then also $V_tU_1^kU_2^l$ and $V_{t'}U_1^kU_2^l$ ($t\not=t'$) are linearly
independent. Since $V_tU_1^kU_2^l$ and $V_{t'}U_1^{k'}U_2^{l'}$
$((k,l)\not=(k',l'))$ shift a different number of steps
in the above basis $e_{kl}$ they are obviously independent. In other words, the
monomials $V_tU_1^kU_2^l$ constitute a basis in $\cO(\T^2)\rtimes\R$. If
$\pi(a)=\sum a_{jkl}V_{t_j}U_1^kU_2^l=0$ we have $a_{jkl}=0$ 
and $\pi$ is a bijection.
\epf\\[1ex]
In analogy with the definition given before Theorem \ref{thm1}, putting
$M=\T^2$ and $\Gamma=\R$, we define the crossed product 
\beq\label{ca}
\cA:=C^\infty(\T^2)\rtimes \R
\eeq
as a
$*$-subalgebra of $B(L^2(\T^2))$.
\bre
We can introduce a set of seminorms on $\cO=\cO(\T^2)\rtimes\R$ as follows.
Let $\cO(\T^2)\subset\cO$ be the $*$-subalgebra
generated by $U_1$ and $U_2$. We define a family $(p_n)_{n\in\N}$ of
seminorms on this subalgebra by
\[
p_n(\sum_{jk}a_{jk}U_1^jU_2^k)=\sup_{j,k\in\Z}(1+|j|+|k|)^n|a_{jk}|.
\]
It is well-known (\cite[22.19.2.-4.]{dieu56}) that the completion of
$\cO(\T^2)$ in the corresponding Fr\'echet topology is $C^\infty(\T^2)$.
Now we define seminorms on $\cO$ by
\[
p_n^\rtimes(\sum_{jkl}a_{jkl}V_{t_j}U_1^kU_2^l)=p_n(\sum_{jkl}a_{jkl}U_1^kU_2^l).
\]
Then it is easy to show that $\cA$ is the completion
$\cO$ with respect to the Fr\'echet topology defined by the family of seminorms
$p_n^\rtimes$. To this end, one first notes that every element of $\cA$ is
a finite sum of products $fV_t$, $f\in C^\infty(\T^2)$. By the above, 
$f$ is a limit of elements $f_k$ of $\cO(\T^2)$ with respect to
$p_n$, and by the definition of $p_n^\rtimes$ it is obvious that $fV_t$ is the
limit of  $f_kV_t$.
\ere

Let us now describe the Hilbert space of the spectral triple of Theorem 
\ref{thm1} for the Kronecker foliation. 

Here, both $V$ and $N$ are one-dimensional, with local
frames consisting each of one vector $\frac{\partial}{\partial x}$
and $\underline{n}=\frac{\partial}{\partial y}+V$ respectively.
Let $\tau$ and $\nu$ denote the corresponding elements of the dual frames.
Then $E=\bigwedge V^*\ot\bigwedge N^*$ consists of four one-dimensional
subspaces of elements of degrees $(0,0),~(1,0),~(0,1),~(1,1)$, with local
frames ${\bf 1}$, $\tau$, $\nu$, $\tau\ot\nu$, respectively. The natural choice of 
translation
invariant (under the natural action of $\R^2$ on $\T^2$) Euclidean fibre
metrics makes these frame elements mutually orthogonal unit vectors
in $L^2(\T^2,E)$. We may identify
\[
L^2(\T^2,E)=L^2(T^2)\oplus L^2(T^2)\oplus L^2(T^2)\oplus L^2(T^2)\enspace,
\]
with $e_{kl}1\ra (e_{kl},0,0,0),\ldots,e_{kl}\tau\ot\nu\ra(0,0,0,e_{kl})$.

\bre
Since the generators act, according to (\ref{gen}), componentwise in $L^2(\T^2,E)$
the crossed product algebra of Theorem \ref{thm1} 
coincides with (\ref{ca}).
\ere
We choose the transversal subspace $H$ in the simplest
way, i.e. we put $h^i_k=0$, i.e. $H$ is generated by the coordinate vector field
$\frac{\partial}{\partial y}$. Then the general formulae
of the foregoing section lead (with some easy computations for the adjoints)
to the following expressions:
{%
\renewcommand{\arraystretch}{2.0}
\[
\begin{array}{c|c|c|c|c}
 & \displaystyle f & \displaystyle f\tau & \displaystyle f\nu & \displaystyle
f\tau\otimes\nu\\
\hline
d_L & \displaystyle\frac{\partial f}{\partial x} & 0 &
\displaystyle\frac{\partial f}{\partial x}\tau\ot\nu & 0\\
d_H &\displaystyle\frac{\partial f}{\partial y}\nu & \displaystyle
-\frac{\partial f}{\partial y}\tau\ot\nu & 0 & 0\\
d_L^* & 0 &\displaystyle-\frac{\partial f}{\partial x} &
0 & \displaystyle-\frac{\partial f}{\partial x}\nu\\
d_H^* &0 &0 &\displaystyle-\frac{\partial f}{\partial y} &
\displaystyle\frac{\partial f}{\partial y}\tau
\end{array}\enspace,
\]
\renewcommand{\arraystretch}{1.0}
}
with $f\in C^\infty(\T^2)$.
To prove, e.g.,
\[
d_H^*(f\tau\otimes\nu)=\frac{\partial f}{\partial y}\tau,
\]
we denote by $(\cdot|\cdot)$ the scalar product in $L^2(\T^2,E)$ and observe that
\begin{eqnarray*}
\left(g\tau|d_H^*(f\tau\otimes\nu)\right)&\equiv&(d_H(g\tau)|f\tau\ot\nu)=(-\frac{\partial
g}{\partial y}\tau\ot\nu|f\tau\ot\nu)\\
&=&-\int\frac{\partial g(x,y)}{\partial y}f(x,y)dxdy
=\int g(x,y)\frac{\partial f(x,y)}{\partial y}
dxdy\\
&=&(g\tau|\frac{\partial f}{\partial y}\tau)\enspace.
\end{eqnarray*}

Note that all the above operators can also be written as matrix differential
operators.

\subsection{The first order signature operator as Dirac operator}

We will first show that $(C^\infty(\T^2)\rtimes\R,L^2(\T^2,E),\tilde{Q})$, with
$\tilde{Q}$ being the linear signature operator
\[
\tilde{Q}=d_L+d_L^*+d_H+d_H^*,
\]
is a spectral triple of dimension 2. Using the foliation chart $(x,y)$
and the local frame $\{{\bf 1},\tau,\nu,\tau\ot\nu\}$, this operator can be written as
\[
\tilde{Q}=\left(\ba{cccc}0&-\frac{\partial}{\partial x}&-\frac{\partial}{\partial y}
&0\\
\frac{\partial}{\partial x}&0&0&\frac{\partial}{\partial y}\\
\frac{\partial}{\partial y}&0&0&-\frac{\partial}{\partial x}\\
0&-\frac{\partial}{\partial y}&\frac{\partial}{\partial x}&0
\ea\right).
\]
The eigenvalue problem for $\tilde{Q}$ is most easily solved by considering 
its square
\[
\tilde{Q}^2=
\left(\ba{cccc}
-\frac{\partial^2}{\partial x^2}-\frac{\partial^2}{\partial y^2}&0&0&0\\
0&-\frac{\partial^2}{\partial x^2}-\frac{\partial^2}{\partial y^2}&0&0\\
0&0&-\frac{\partial^2}{\partial x^2}-\frac{\partial^2}{\partial y^2}&0\\
0&0&0&-\frac{\partial^2}{\partial x^2}-\frac{\partial^2}{\partial y^2}
\ea\right).
\]
In coordinates $(\vartheta_1,\vartheta_2)$ one is immediately led to the 
eigenvalue equations
\[
\left (-\left (a\frac{\partial}{\partial \vartheta_1}
+b\frac{\partial}{\partial \vartheta_2}\right )^2
-\left(b\frac{\partial}{\partial \vartheta_1}
-a\frac{\partial}{\partial \vartheta_2}\right)^2\right)f_j=\lambda^2 f_j\enspace,
\]
for the four components of an eigenvector $f=f_{\bf 1}+f_2\tau+f_3\nu+f_4\tau\ot\nu$ of
$\tilde{Q}^2$. It is now straightforward to see that 
\[
e^1_{kl}=\left(\ba{c} e_{kl}\\0\\0\\0\ea\right),
e^2_{kl}=\left(\ba{c} 0\\e_{kl}\\0\\0\ea\right),
e^3_{kl}=\left(\ba{c} 0\\0\\e_{kl}\\0\ea\right),
e^4_{kl}=\left(\ba{c} 0\\0\\0\\e_{kl}\ea\right)
\]
are eigenvectors of $\tilde{Q}^2$ to the eigenvalue 
\[
\lambda^{2}_{kl}=(ak+bl)^2+(al-bk)^2.
\]
The operator $\tilde{Q}$ itself acts in this basis of $L^2(\T^2,E)$ as follows:
\begin{eqnarray*}
\tilde{Q}(e^1_{kl})&=&(ak+bl)e^2_{kl}+(al-bk)e^3_{kl}\\
\tilde{Q}(e^2_{kl})&=&(-ak-bl)e^1_{kl}+(-al+bk)e^4_{kl}\\
\tilde{Q}(e^3_{kl})&=&(-al+bk)e^1_{kl}+(ak+bl)e^4_{kl}\\
\tilde{Q}(e^4_{kl})&=&(al-bk)e^2_{kl}-(ak+bl)e^3_{kl}\enspace.
\end{eqnarray*}
Thus, we have 
\beq\label{qeva}
\lambda^{\pm}_{kl}=\pm\sqrt{(ak+bl)^2+(al-bk)^2}
\eeq
and 
\begin{eqnarray*}
e^{+1}_{kl}&=&-i\frac{bk-al}{\lambda^{+}_{kl}}e^1_{kl}+e^3_{kl}+
i\frac{ak+bl}{\lambda^{+}_{kl}}e^4_{kl},\\
e^{+2}_{kl}&=&-i\frac{ak+bl}{\lambda^{+}_{kl}}e^1_{kl}+e^2_{kl}+
i\frac{al-bk}{\lambda^{+}_{kl}}e^4_{kl},\\
e^{-1}_{kl}&=&e^1_{kl}-i\frac{ak+bl}{\lambda^{+}_{kl}}e^2_{kl}+
i\frac{al-bk}{\lambda^{+}_{kl}}e^3_{kl},\\
e^{-2}_{kl}&=&i\frac{al-bk}{\lambda^{+}_{kl}}e^2_{kl}+         
i\frac{ak+bl}{\lambda^{+}_{kl}}e^3_{kl}+e^4_{kl}
\end{eqnarray*}
form a complete set of eigenvectors of $\tilde{Q}$.
\bpr\label{spec}
$(C^\infty(\T^2)\rtimes\R,L^2(\T^2,E),\tilde{Q})$ is a spectral triple of
dimension 2. \epr
\bpf The eigenvalues (\ref{qeva}) of $\tilde{Q}$ have finite multiplicity, 
tend to infinity for $k,l\to\infty$ and have no
finite accumulation point. Thus, the resolvent of $\tilde{Q}$ is compact.
Since
\beq\label{qvt}
[\tilde{Q},V_t]=0,
\eeq
boundedness of the commutators of $\tilde{Q}$ with
elements of the algebra follows from the fact that
\[
[\tilde{Q},fV_t]=f[\tilde{Q},V_t]+[\tilde{Q},f]V_t=
\left(\ba{cccc}0&-\frac{\partial f}{\partial x}&-\frac{\partial f}{\partial y}
&0\\
\frac{\partial f}{\partial x}&0&0&\frac{\partial f}{\partial y}\\
\frac{\partial f}{\partial y}&0&0&-\frac{\partial f}{\partial x}\\
0&-\frac{\partial f}{\partial y}&\frac{\partial f}{\partial x}&0
\ea\right)V_t,
\]
 is a bounded matrix multiplication operator in 
$L^2(\T^2,E)$.
In order to see that the $n$-th eigenvalue of $|\tilde{Q}|$ is of order $\sqrt{n}$
notice first that the eigenvalues of $|\tilde{Q}|$ are $\lambda^+_{kl}$, with
multiplicity $4\times $(number of $(k,l)\in \Z^2$ leading to the
same $\lambda^+_{kl}$). The number of eigenvalues with absolute value less
than some $R>0$ is then $4\times$(number of integer lattice points
inside a circle of radius $R$), i.e. equal to $4\times$(the area $\pi R^2$)
up to lower
order terms in $R$. (Recall that $(x,y)\mapsto(\vartheta_1,\vartheta_2)$ 
is orthogonal.) This proves the claim.
\epf\\[1ex]
Note that the commutators of $\tilde{Q}$ with the generators $U_1$ and $U_2$
are explicitly given by
\begin{eqnarray*}
\left[{\tilde{Q}},U_1\right]e^1_{kl}&=&ae^2_{k+1,l}-be^3_{k+1,l},\\
\left[{\tilde{Q}},U_1\right]e^2_{kl}&=&-ae^1_{k+1,l}+be^4_{k+1,l},\\
\left[{\tilde{Q}},U_1\right]e^3_{kl}&=&be^1_{k+1,l}+ae^4_{k+1,l},\\
\left[{\tilde{Q}},U_1\right]e^4_{kl}&=&-be^2_{k+1,l}-ae^3_{k+1,l}
\end{eqnarray*}
and 
\begin{eqnarray*}
\left[\tilde{Q},U_2\right]e^1_{kl}&=&be^2_{k,l+1}+ae^3_{k,l+1},\\
\left[\tilde{Q},U_2\right]e^2_{kl}&=&-be^1_{k,l+1}-ae^4_{k,l+1},\\
\left[\tilde{Q},U_2\right]e^3_{kl}&=&-ae^1_{k,l+1}+be^4_{k,l+1},\\
\left[\tilde{Q},U_2\right]e^4_{kl}&=&ae^2_{k,l+1}-be^3_{k,l+1}.
\end{eqnarray*}

In order to describe the differential algebra
$\Omega_{\tilde{Q}}(\cO(\T^2)\rtimes\R)$, we denote, as in formulae
(\ref{o1}) and (\ref{o2}), by $\pi^1$ and $\pi^2$ 
the extensions
of $\pi$ to universal one and two forms. Since $\pi$ is faithful
by Proposition \ref{alg},
$\Omega^1_{\tilde{Q}}(\cO(\T^2)\rtimes\R)$ is 
isomorphic to $\pi^1(\Omega^1(\cO(\T^2)\rtimes\R))$, with $du_j\mapsto
[\tilde{Q},U_j]$, $dv_t\mapsto [\tilde{Q},V_t]$, 
and 
$\Omega^2_{\tilde{Q}}(\cO(\T^2)\rtimes \R)=
\Omega^2(\cO(\T^2)\rtimes \R)/(\ker\pi^2+d(\ker\pi^1))\simeq
\pi^2(\Omega^2(\cO(\T^2)\rtimes \R))/\pi^2(d(\ker\pi^1))$.

Let us first note that, under the
identification $L^2(\T^2,E)\simeq \C^4\ot L^2(\T^2)$ given by
\beq
e^1_{kl}\mapsto \left(\ba{c}1\\0\\0\\0\ea\right)\ot e_{kl},~
e^2_{kl}\mapsto \left(\ba{c}0\\1\\0\\0\ea\right)\ot e_{kl},~
e^3_{kl}\mapsto \left(\ba{c}0\\0\\1\\0\ea\right)\ot e_{kl},~
e^4_{kl}\mapsto \left(\ba{c}0\\0\\0\\1\ea\right)\ot e_{kl},
\eeq
$U_1,U_2,V_t$ and the above commutators can be written as follows:
\beq\label{mat}
U_1=\id\ot s_1,~U_2=\id\ot s_2,~V_t=\id\ot v_{abt},
\eeq
where $s_1e_{kl}=e_{k+1,l}$, $s_2e_{kl}=e_{k,l+1}$,
$v_{abt}e_{kl}={\rm e}^{{\rm i}(ak+bl)t}e_{kl}$, and
\beq\label{mac}
[\tilde{Q},U_1]=\left(\ba{rrrr}0&a&-b&0\\
-a&0&0&b\\
b&0&0&a\\
0&-b&-a&0\ea\right)\ot s_1,~~
[\tilde{Q},U_2]=\left(\ba{rrrr}
0&b&a&0\\
-b&0&0&-a\\
-a&0&0&b\\
0&a&-b&0\ea\right)\ot s_2.
\eeq
Using this representation, together with $[s_1,s_2]=0$, 
$s_1v_{abt}=e^{iat}v_{abt}s_1$, $s_2v_{abt}=e^{ibt}v_{abt}s_2$,
it is easy to show
\ble
\beq\label{ujk}
U_j[\tilde{Q},U_k]=[\tilde{Q},U_k]U_j,~~\forall j,k\in\{1,2\},
\eeq
\beq\label{uvt}
V_t[\tilde{Q},U_1]=e^{iat}[\tilde{Q},U_1]V_t,~~V_t[\tilde{Q},U_2]=e^{ibt}[\tilde{Q},U_2]V_t,
\eeq
\beq\label{qu12}
[{\tilde{Q}},U_1][\tilde{Q},U_2]=-[\tilde{Q},U_2][\tilde{Q},U_1].
\eeq
\ele
Explicitly, we have
\beq\label{qu1u2}
[\tilde{Q},U_1][\tilde{Q},U_2]=\left(\ba{rrrr}
0&0&0&-1\\0&0&-1&0\\0&1&0&0\\1&0&0&0
\ea\right)\ot s_1s_2\enspace.
\eeq
\bpr\label{oql}
\begin{enumerate}
\item[(i)] $\Omega^1_{\tilde{Q}}(\cO(\T^2)\rtimes\R)$ is a free left (and right)
$\cO(\T^2)\rtimes\R$-module with basis $\{du_1,du_2\}$. Its bimodule
structure is determined by
\beq\label{dujk}
u_jdu_k=du_ku_j,~~\forall j,k\in\{1,2\},
\eeq
\beq\label{duvt}
v_tdu_1=e^{iat}du_1v_t,~~v_tdu_2=e^{ibt}du_2v_t.
\eeq
Moreover,
\beq\label{dvt}
dv_t=0.
\eeq
\item[(ii)] $\Omega^2_{\tilde{Q}}(\cO(\T^2)\rtimes\R)$ is a free left (and right)
$\cO(\T^2)\rtimes\R$-module with basis $\{du_1du_2\}$, with
\beq\label{du12}
du_1du_2=-du_2du_1.
\eeq
\item[(iii)] $\Omega^k_{\tilde{Q}}(\cO(\T^2)\rtimes\R)=0$ for $k\geq 3$.
\end{enumerate}
\epr
\bpf 
Since the representation 
$\pi$ of $\cO(\T^2)\rtimes\R$ in $L^2(\T^2,E)$ is faithful,
the equations (\ref{dujk}),
(\ref{duvt}) and (\ref{du12}) follow from (\ref{ujk})--(\ref{qu12}),
whereas (\ref{dvt}) comes from (\ref{qvt}). Now it is sufficient to show
that every element of $\pi^1(\Omega^1(\cO(\T^2)\rtimes\R))$ is of the
form $a_1[\tilde{Q},U_1]+a_2[\tilde{Q},U_2],~~a_1,a_2\in \pi(\Omega^1(\cO(\T^2)\rtimes\R))$ 
and that from $a_1[\tilde{Q},U_1]+a_2[\tilde{Q},U_2]=0$ follows $a_1=a_2=0$. The first
claim is immediate from the fact that $\pi(\cO(\T^2)\rtimes\R)$ is generated
by $U_1,U_1^*,U_2,U_2^*,V_t$ and from (\ref{qvt}),
(\ref{ujk})--(\ref{qu12}). (Note that commutators $[\tilde{Q},U_j^*]$ can be reduced
to $[\tilde{Q},U_j]$ using the Leibniz rule: From $0=[\tilde{Q},U_j^*U_j]=U_j^*[\tilde{Q},U_j]+
[\tilde{Q},U_j^*]U_j$ follows $[\tilde{Q},U_j^*]=-U_j^*[\tilde{Q},U_j]U_j^*$. On the other hand,
multiplying $[\tilde{Q},U_j]U_j=U_j[\tilde{Q},U_j]$ from both sides with $U_j^*$ gives
$U_j^*[\tilde{Q},U_j]=[\tilde{Q},U_j]U_j^*$.) It remains to show linear independence.
Assume 
\[
\sum\limits_{ij}\left( a_{ij}V_{t_{ij}}U_1^iU_2^j[\tilde{Q},U_1]+
b_{ij}V_{t'_{ij}}U_1^iU_2^j[\tilde{Q},U_2]\right)=0\enspace,
\]
$a_{ij},b_{ij}\in\C$, finite summation over $i,j\in\Z$.
Acting with this expression on the basis vector $e^1_{kl}$ gives 
\begin{eqnarray*}
&&a_{ij}e^{it_{ij}((k+i+1)a+(l+j)b)}(ae^2_{k+i+1,l+j}-be^3_{k+i+1,l+j})\\
&&=
b_{i+1,j-1}e^{it'_{i+1,j-1}((k+i+1)a+(l+j)b)}(be^2_{k+i+1,l+j}+
ae^3_{k+i+1,l+j}),
\end{eqnarray*}
(now for fixed $i,j$), which means
\begin{eqnarray*}
aa_{ij}e^{it_{ij}((k+i+1)a+(l+j)b)}-
bb_{i+1,j-1}e^{it'_{i+1,j-1}((k+i+1)a+(l+j)b)}&=&0\\
ba_{ij}e^{it_{ij}((k+i+1)a+(l+j)b)}+
ab_{i+1,j-1}e^{it'_{i+1,j-1}((k+i+1)a+(l+j)b)}&=&0\enspace.
\end{eqnarray*}
Since 
\begin{eqnarray*}
&\det\left(\ba{rr}ae^{it_{ij}((k+i+1)a+(l+j)b)}&-be^{it'_{i+1,j-1}((k+i+1)a+(l+j)b)}\\
be^{it_{ij}((k+i+1)a+(l+j)b)}&ae^{it'_{i+1,j-1}((k+i+1)a+(l+j)b)}\ea\right)
=&\\
&=(a^2+b^2)e^{i(t_{ij}+t'_{i+1,j-1})((k+i+1)a+(l+j)b)}
=
e^{i(t_{ij}+t'_{i+1,j-1})((k+i+1)a+(l+j)b)}\neq 0\enspace,&
\end{eqnarray*}
this
system has only the trivial solution $a_{ij}=b_{i+1,j-1}=0$. Thus (i) is
proven.\\
To prove (ii) note that differentiating (\ref{dujk}) for $j=k$ leads
immediately to $du_1du_1=du_2du_2=0$. Moreover, (\ref{qu12}) yields
$du_1du_2=-du_2du_1$, so that $\Omega^2_{\tilde{Q}}(\cO(\T^2)\rtimes \R)$ is generated
by $du_1du_2$. It remains to show that it is freely generated. 
To this end we have to determine $\ker\pi^1$.
From (\ref{ujk}) and (\ref{uvt}) it follows that $\ker\pi^1$ 
contains the bimodule generated by the elements $u_jdu_k-du_ku_j$,
$v_tdu_1-e^{iat}du_1v_t$, $v_tdu_2-e^{ibt}du_2v_t$. On the other hand,
this bimodule also contains $\ker \pi^1$:
Let
\[
\alpha=\sum a^r_{t_jkl,t_mnq}v_{t_j}u_1^ku_2^ldu_rv_{t_m}u_1^nu_2^q\in
\ker\pi^1. 
\]
Using the commutation rules (\ref{ujk}) and (\ref{uvt}) and the basis property
of the $[\tilde{Q},u_i]$ already proved in (i), one concludes from
$\pi^1(\alpha)=0$ that
\[
\sum a^r_{t_jkl,t_mnq}v_{t_j+t_m}u_1^{k+n}u_2^{l+q}e^{-i((k+1)a+lb)t_m}=0, 
~~r=1,2.
\]
Now, making use of the basis property of the monomials $v_tu_1^ku_2^l$
(Proposition \ref{alg}), one has
\beq\label{arj}
\sum_{t_j+t_m=T,k+n=K,l+q=L}a^r_{jkl,mnq}e^{-i((k+1)a+lb)t_m}=0, ~~r=1,2,
\eeq
for every fixed $T,K,L$. Now fix (for $r=1$) $k_0,l_0,t_{j_0}$
and write equation (\ref{arj}) as
\beas
a^1_{t_{j_0}k_0l_0,T-t_{j_0},K-k_0,L-l_0}=&&\\[.3cm]
-e^{i((k_0+1)a+l_0b)(T-t_{j_0})}
&\sum_{t_j\neq t_{j_0},k\neq k_0,l\neq l_0}\limits&
a^1_{t_jkl,T-t_j,K-k,L-l}e^{-i((k+1)a+lb)(T-t_j)}.
\eeas
Inserting this into $\alpha$ one obtains\\[.5cm]
$\alpha=\sum_{(t_j,k,l)\neq(t_{j_0},k_0,l_0)}\limits
a^1_{t_jkl,T-t_j,K-k,L-l}v_{t_j}u_1^ku_2^ldu_1v_{T-t_j}u_1^{K-k}u_2^{L-l}
\\[.4cm]
-e^{i((k_0+1)a+l_0b)(T-t_{j_0})}
\sum_{(t_j,k,l)\neq(t_{j_0},k_0,l_0)}\limits
a^1_{t_jkl,T-t_j,K-k,L-l}v_{t_{j_0}}u_1^{k_0}u_2^{l_0}du_1v_{T-t_{j_0}}
u_1^{K-{k_0}}u_2^{L-{l_0}}\\[.3cm]
\hspace*{.5cm}\cdot e^{i((k+1)a+lb)(T-t_{j})}+\mbox{(a similar term for
$r=2$)}.
$\\[.5cm]
Now, fix $t_j>t_{j_0},k>k_0,l>l_0$, and reduce the power of $u_2$ in front of
$du_1$ in the first term by subtracting and adding $du_1u_2$, thus producing
a term in the bimodule (with $u_2du_1-du_1u_2$ in the middle) and a new term
with a new $e$-factor
of the old kind. Iterating this procedure removes all superfluous powers
of $u_2$. One can do the same for $u_1$ and $v_t$ and finally ends up with 
an expression which turns out to be zero (up to a lot of terms all lying
in the bimodule). We leave the detailed computation to the reader.
Thus we have shown that $\ker\pi^1$ is also contained in the bimodule generated by the
elements (\ref{dujk}) and (\ref{duvt}).

Therefore, a general element of $\ker\pi^1$ is of the form 
\[
j=\sum a_k\alpha_k b_k
\]
with $a_k,b_k\in \cO(\T^2)\rtimes \R$, $\alpha_k$ one of the above
generating elements of $\ker\pi^1$. Then
\[
\pi^2(dj)=\sum\pi(a_k)\pi^2(d\alpha_k)\pi(b_k),
\]
because the other terms appearing according to the Leibniz rule contain a
factor $\pi^1(\alpha_k)=0$. We have to determine $\pi^2\ci
d(u_jdu_k-du_ku_j)=\pi^2(du_jdu_k+du_kdu_j)=[\tilde{Q},U_j][\tilde{Q},U_k]+
[\tilde{Q},U_k][\tilde{Q},U_j]$.
A trivial calculation using (\ref{mac}) and
(\ref{mat}) shows that $[\tilde{Q},U_j][\tilde{Q},U_j]=-U_j^2$, whereas
$[\tilde{Q},U_1][\tilde{Q},U_2]+[\tilde{Q},U_2][\tilde{Q},U_1]=0$ by (\ref{qu12}). It follows that
\[
\pi^2(d\ker\pi^1)=\pi(\cO(\T^2)\rtimes \R),
\]
and it remains to show that from $\pi(a)[\tilde{Q},U_1][\tilde{Q},U_2]\in
\pi(\cO(\T^2)\rtimes \R)$ follows $a=0$. Indeed, this follows from the fact
that algebra elements have the diagonal form (\ref{mat}) whereas
$[\tilde{Q},U_1][\tilde{Q},U_2]$ is antidiagonal (\ref{qu1u2}), which is preserved under
multiplication with a diagonal element $\pi(a)$.\\
(iii) is a trivial consequence of the fact that in a form of degree $\geq 3$
at least two differentials of the same generator $u_j$ will meet to produce
0.
\epf
\bre\label{topcal}
One can define a first order differential calculus for the algebra
$\cA=C^\infty(\T^2)\rtimes \R$ in the following way:
Let $\Omega^1_{\tilde{Q}}(\cA)$ be the free left $\cA$-module with basis
$\{du_1,du_2\}$. Equipped with the product of the topologies defined by
the sequence of seminorms $p_n^\rtimes$, 
$\Omega^1_{\tilde{Q}}(\cA)$ is a free left topological $\cA$-module. One
turns
$\Omega^1_{\tilde{Q}}(\cA)$ into a right $\cA$-module by defining
$a_jdu_ju_k:=a_ju_kdu_j$ for $j,k\in{1,2}$, $a_1du_1v_t:=e^{iat}a_1v_tdu_1$ and 
$a_2du_2v_t:=e^{ibt}a_2v_tdu_2$ ($a_j\in \cA$), and extending this by
continuity. This gives $\Omega^1_{\tilde{Q}}(\cA)$ 
the structure of a topological bimodule containing
$\Omega^1_{\tilde{Q}}(\cO)$ as a dense subbimodule. It is also not
difficult to see that the
differential can be extended to a continuous map 
$d:\cA\lra \Omega^1_{\tilde{Q}}(\cA)$. Analogously, one can define
a topological $\cA$-bimodule $\Omega^1_{\tilde{Q}}(\cA)$ such that the
natural mappings $\Omega^1_{\tilde{Q}}(\cA)\times\Omega^1_{\tilde{Q}}(\cA)
\lra \Omega^2_{\tilde{Q}}(\cA)$ and $d:\Omega^1_{\tilde{Q}}\lra\Omega^2_{\tilde{Q}}(\cA)$
are continuous. We conjecture that the differential calculus
$\Omega_{\tilde{Q}}(\cA)$ so constructed coincides with the calculus
(to be denoted by the same symbol) resulting from the spectral triple
$(\cA,L^2(\T^2,E),\tilde{Q})$.
\ere

\subsection{The mixed signature operator}

Let us now consider the mixed signature operator $Q$ given by formula
(\ref{qmix}). In matrix representation, we have
\[
Q=\left(
\ba{cccc}
\frac{\partial ^2}{\partial x^2}&0&\frac{\partial }{\partial y}&0\\
0&-\frac{\partial^2 }{\partial x^2}&0&-\frac{\partial}{\partial y}\\
-\frac{\partial}{\partial y}&0&-\frac{\partial^2 }{\partial x^2}&0\\
0&\frac{\partial}{\partial y}&0&\frac{\partial^2 }{\partial x^2}\\
\ea
\right).
\]
In order to diagonalize this operator, we have to solve the eigenvalue
problem
\beq\label{eigenv}
Q
\left(
\ba{c}
f_1\\f_2\\f_3\\f_4
\ea\right)=
\lambda
\left(
\ba{c}
f_1\\f_2\\f_3\\f_4
\ea\right)
\eeq
with $f_i\in L^2(\T^2,E)$. $Q$ is already block-diagonal and acts in the same way
in the space of $(0,0)$- and $(0,1)$-forms and in the space
of $(1,1)$- and $(1,0)$-forms. It suffices to diagonalize one block.
Defining
\begin{eqnarray*}
g&=&f_1+f_3\\
h&=&f_1-f_3,
\end{eqnarray*}
one arrives at 
\begin{eqnarray*}
\frac{\partial^2 h }{\partial x^2}+\frac{\partial h}{\partial y}
&=&\lambda g\\
\frac{\partial^2g }{\partial x^2}-\frac{\partial g }{\partial y}
&=&\lambda h\enspace.
\end{eqnarray*}
Returning to the original coordinates $(\vartheta_1,\vartheta_2)$,
the foregoing equations read
\beq\label{h}
a^2\frac{\partial^2h}{\partial \vartheta_1^2}+2ab\frac{\partial^2h}{\partial
\vartheta_1\partial
\vartheta_2}+b^2\frac{\partial^2h}{\partial \vartheta_2^2}-b\frac{\partial h}{\partial
\vartheta_1}
+a\frac{\partial h}{\partial \vartheta_2}=\lambda g,
\eeq
\beq\label{g}
a^2\frac{\partial^2g}{\partial \vartheta_1^2}+2ab\frac{\partial^2g}{\partial
\vartheta_1\partial
\vartheta_2}+b^2\frac{\partial^2g}{\partial \vartheta_2^2}+b\frac{\partial g}{\partial
\vartheta_1}
-a\frac{\partial g}{\partial \vartheta_2}=\lambda h.
\eeq
The ansatz 
\begin{eqnarray*}
g&=&\sum_{k,l\in\Z}\eta_{kl}e^{i(k\vartheta_1+l\vartheta_2)}\\
h&=&\sum_{k,l\in\Z}\chi_{kl}e^{i(k\vartheta_1+l\vartheta_2)}
\end{eqnarray*}
leads to
\[
\left(\left(a^2k^2+2abkl+b^2l^2\right)^2+(bk-al)^2\right)\chi_{kl}=
\lambda^2\chi_{kl},
\]
which gives the eigenvalues
\[
\lambda_{kl\pm}=\pm\sqrt{(ak+bl)^4+(bk-al)^2}.
\]
One easily concludes that eigenvectors to the eigenvalues $\lambda_{kl\pm}$
are of the form
\[
h_{kl}=e_{kl}, ~~g_{kl\pm}=\gamma_{kl\pm}e_{kl}
\]
with
\[
\gamma_{kl\pm}=\frac{-(ak+bl)^2+i(al-bk)}{\lambda_{kl\pm}}.
\]
The eigenvectors of the original problem (\ref{eigenv}) are
\begin{eqnarray*}
{f_1}_{kl\pm}&=&\frac{1}{2}\left(g_{kl\pm}+h_{kl}\right)=
\frac{1}{2}\left(1+\gamma_{kl\pm}\right)e_{kl}\nonumber\\
{f_3}_{kl\pm}&=&\frac{1}{2}\left(g_{kl\pm}-h_{kl}\right)=
\frac{1}{2}\left(\gamma_{kl\pm}-1\right)e_{kl}, 
\end{eqnarray*}
or, written as elements of $L^2(\T^2,E)$,
\[
e^{(1)}_{kl\pm}=\frac{1}{2}e_{kl}\left((\gamma_{kl\pm}+1){\bf 1}
+(\gamma_{kl\pm}-1)\nu\right).
\]
If we assume that the metrics are chosen so that the frame elements
${\bf 1},\tau,\nu,\tau\ot\nu$ are orthonormal, these vectors are already orthonormal
(note that $|\gamma_{kl\pm}|=1$.)
The same argument yields another set
\[
e^{(2)}_{kl\pm}=\frac{1}{2}e_{kl}\left((\gamma_{kl\pm}+1)\tau\ot\nu+
(\gamma_{kl\pm}-1)\tau\right)
\]
of eigenvectors to the same eigenvalues $\lambda_{kl\pm}$.
Note that the eigenvalue $0$ appears only for $k=l=0$.
In that case, equations (\ref{h}) and (\ref{g}) decouple, and we get
four independent eigenvectors ${\bf 1},\tau,\nu,\tau\ot\nu$. In order to see that
these vectors together with the $e^{(1,2)}_{kl\pm}$ form
an orthonormal basis of $L^2(\T^2,E)$, it is sufficient to see that
all the vectors $e_{kl}{\bf 1},~e_{kl}\tau,~~e_{kl}\nu,~~e_{kl}\tau\ot\nu$
are linear combinations of the foregoing vectors.
This follows from the fact that the matrix
\[
\left(\ba{cc}
\gamma_{kl+}+1&\gamma_{kl+}-1\\
\gamma_{kl-}+1&\gamma_{kl-}-1\\
\ea\right)
\]
is always invertible (its determinant being $-4\gamma_{kl+}$).\\
Thus we have found the spectral decomposition of the selfadjoint
operator $Q$. Its unboundedness is reflected in the unboundedness
of the $\lambda_{kl\pm}$. It is now easy to write down also 
the spectral decomposition of the corresponding Dirac operator $D$:
Applying (\ref{Q}) for nonzero eigenvalues gives
\[
De^{(1,2)}_{kl\pm}=\pm\sqrt{\lambda_{kl}}e^{(1,2)}_{kl\pm},
\]
where $\lambda_{kl}$ is the positive root $\lambda_{kl+}$.
Putting $e^{(1)}_{00+}={\bf 1},e^{(1)}_{00-}=\nu,e^{(2)}_{00+}=\tau\ot\nu$ and
$e^{(2)}_{00-}=\tau$, the formula defines $D$ also on the kernel of $Q$
(cf. \ref{d|d|}), and gives the spectral decomposition of $D$.

The unitary operators $U_1$, $U_2$ and $V_t$  act by (\ref{gena}) 
on the basis vectors
$e_{kl\pm}^{(1,2)}$ as follows
\begin{eqnarray*}
U_1e^{(1,2)}_{kl\pm}&=&\frac{1}{2}\left\{\left(1+
\frac{\gamma_{kl\pm}}{\gamma_{k+1,l+}}
\right) e^{(1,2)}_{k+1,l+}+\left(1+\frac{\gamma_{kl\pm}}{\gamma_{k+1,l-}}\right)
e^{(1,2)}_{k+1,l-}\right\}\\
U_2e^{(1,2)}_{kl\pm}&=&\frac{1}{2}\left\{\left(1+
\frac{\gamma_{kl\pm}}{\gamma_{k,l+1,+}}\right)
e^{(1,2)}_{k,l+1,+}+\left(1+\frac{\gamma_{kl\pm}}{\gamma_{k,l+1,-}}\right)
e^{(1,2)}_{k,l+1,-}\right\}\\
V_te^{(1,2)}_{kl\pm}&=&{\rm e}^{{\rm i}(ka+lb)t}e^{(1,2)}_{kl\pm}\enspace.
\end{eqnarray*}
Defining
\[
\eta^{(1,2)}_{kl\pm}:=\frac{1}{2}\left(e^{(1,2)}_{kl+}\pm e^{(1,2)}_{kl-}\right)\enspace,
\]
one finds
\begin{eqnarray}\label{u1e+}
U_1\eta^{(1,2)}_{kl+}&=&\eta^{(1,2)}_{k+1,l+}\enspace,\\
\label{u1e-}
U_1\eta^{(1,2)}_{kl-}&=&\frac{\gamma_{kl}}{\gamma_{k+1,l}}\eta^{(1,2)}_{k+1,l-}\enspace,\\
\label{u2e+}
U_2\eta^{(1,2)}_{kl+}&=&\eta^{(1,2)}_{k,l+1,+}\enspace,\\
\label{u2e-}
U_2\eta^{(1,2)}_{kl-}&=&\frac{\gamma_{kl}}{\gamma_{k,l+1}}\eta^{(1,2)}_{k,l+1,-}\enspace,\\
\label{ve}
V_t\eta^{(1,2)}_{kl\pm}&=&e^{i(ka+lb)t}\eta^{(1,2)}_{kl\pm}\enspace,\\
\label{de}
D\eta^{(1,2)}_{kl\pm}&=&\sqrt{\lambda_{kl}}\eta^{(1,2)}_{kl\mp}\enspace.
\end{eqnarray}
From Theorem~\ref{thm1} or by direct computation using
(\ref{u1e+})--(\ref{de}) one gets
\bpr 
$(C^\infty(\T^2)\rtimes \R,L^2(\T^2,E),D)$ is a spectral triple of dimension 3.
\epr

Next, one would like to describe the differential calculus $\Omega_D$
related to this spectral triple. Unfortunately, we have no definite result
about $\Omega_D$. We will however show that the first order calculus for 
the restriction
of the spectral triple to the subalgebra $C^\infty(\T^2)$ is the universal
one, supporting our conjecture that also the first order calculus of the
full triple is universal, up to some relations involving $V_t$. 
To begin with, we have   
\ble\label{lem2}
Let $p,q,r,s\in\Z$. Then we have 
\begin{eqnarray*}
U_1^rU_2^p\left[D,U_1^sU_2^q\right]\eta_{kl\pm}^{(1,2)}&=&
\frac{\sqrt{\lambda_{k+s,l+q}}\gamma_{k+s,l+q}-
\sqrt{\lambda_{kl}}\gamma_{kl}}{\gamma_{k+r+s,l+p+q}}
\eta_{k+r+s,l+p+q\mp}^{(1,2)}\\
\left[D,V_t\right]&=&0\enspace.
\end{eqnarray*}
Moreover,
\beq\label{vdu}
V_t[D,U_1]={\rm e}^{{\rm i}at}[D,U_1]V_t,~~V_t[D,U_2]={\rm e}^{{\rm i}bt}[D,U_2]V_t.
\eeq
\ele
\bpf By direct computation using (\ref{u1e+})--(\ref{de}).\epf\\[1ex]
From Theorem \ref{thm1} we know that the particular choice 
$\Gamma={\id}$
gives rise to a spectral triple over $C^\infty(\T^2)$. Let us now first 
investigate 
the corresponding differential calculus $\Omega_D({\cal O}(\T^2))$. 
By faithfulness of the representation, we can again identify
$\Omega^1_D({\cal O}(\T^2))$ with a subspace of $B(L^2(\cH))$. We have
\bpr 
The first order differential calculus $\Omega_D^1({\cal O}(\T^2))$ is 
freely generated by the elements $[D,U_1^sU_2^q]$ ($s,q\in\Z$). 
\epr
\bpf 
We show that no nontrivial relations between $U_1,U_2$ and $D$ can exist.
Let us first consider relations between $D$ and $U_1$ only. From Lemma
\ref{lem2} it follows for $p=q=0$ that
\[
U_1^r\left[D,U_1^s\right]\eta_{kl\pm}^{(1,2)}=
\frac{\sqrt{\lambda_{k+s,l}}\gamma_{k+s,k}-
\sqrt{\lambda_{kl}}\gamma_{kl}}{\gamma_{k+r+s,k}}\eta_{k+r+s,k\mp}^{(1,2)}\enspace.
\]
Using the Leibniz rule and the fact that different overall powers of $U_1$
are independent we find that nontrivial relations would be of the form
\begin{equation}\label{3.83}
\sum\limits_{m=0}^{s-1}a_mU_1^m\left[D,U_1^{s-m}\right]=0\enspace,
\end{equation}
for $s\in\N$. Applying (\ref{3.83}) to $\eta_{n0\pm}^{(1,2)}$
($n=k,\dots,k+s-1)$ we get the following system of equations
\begin{eqnarray*}
\sum\limits_{j=0}^{s-1}
a_j\left(\sqrt{\lambda_{k+j+1,0}}\gamma_{k+j+1,0}-\sqrt{\lambda_{k0}}
\gamma_{k0}\right)&=&0\\
&\vdots &\\
\sum\limits_{j=0}^{s-1}a_j\left(\sqrt{\lambda_{k+j+s,0}}\gamma_{k+j+s,0}-
\sqrt{\lambda_{k+s-1,0}}\gamma_{k+s-1,0}\right)&=&0\enspace.
\end{eqnarray*}
For the discussion of this system of equations it is useful to define a
function $h$ on $\Z$ putting
\[
h(i)=\sqrt{\lambda_{i0}}\gamma_{i0}-\sqrt{\lambda_{i-1,0}}
\gamma_{i-1,0}\enspace.
\]
\ble\label{lem3}
We have 
\[
\left|
\begin{array}{ccc}
h(i_0) & \cdots & h(i_0+k)\\
\vdots & \ddots & \vdots\\
h(i_k) & \cdots & h(i_k+k)
\end{array}
\right|\not= 0\enspace,
\]
for all $k\in\N$ and $i_0,\dots,i_k\in\Z$. 
\ele
\bpf
See appendix~\ref{appa}.
\epf\\[1ex]
Thus, there are no relations between $U_1$ and $D$ besides
the ones coming from the Leibniz rule. 
In the general case we are looking for $a_{mn}\in\C$ such that
\[
\sum\limits_{m=0}^{s-1}\sum\limits_{n=0}^{q-1}a_{mn}U_1^mU_2^n\left[D,U_1^{s-m}
U_2^{q-n}\right]=0\enspace.
\]

Again, we are led to the consideration of a homogeneous linear system of 
equations for the $a_{mn}$. The corresponding matrix of coefficients is an
$(sq)\times(sq)$-matrix with general matrix element
\[
C_{k,(m,n)}=\left(\sqrt{\lambda_{k+s-m,q-n}}\gamma_{k+s-m,q-n}-
\sqrt{\lambda_{k0}}\gamma_{k0}\right)\enspace,
\]
($k=1,\dots,sq$.) In analogy to the case discussed above we have 
\ble\label{lem4}
Let $s,q\in\N$ be fixed. Then we have
\[
\det\left(C_{k,(m,n)}\right)\not=0\enspace.
\]
\ele
\bpf
The proof is a straightforward generalization of the proof of Lemma \ref{lem3}
to the case of functions, defined on $\Z^2$, see~\cite{ri1}.
\epf\\[1ex]
The proof of the proposition follows now immediately from the fact that between
the elements $[D,U_1^sU_2^q]$ there are no relations besides the ones coming
from the Leibniz rule.
\epf\\[1ex]
We were not able to derive more relations of the type
(\ref{vdu}) between commutators of $D$ with some generator 
and other generators (up to such relations resulting from applying 
$[D,\cdot]$ to (\ref{u1u2})--(\ref{vv})  and the unitarity condition, using the
derivation property). This seems to be due to the fact that
$\lambda_{kl}$ and $\gamma_{kl}$ contain second and fourth powers of
$k$ and $l$ under the square root. Therefore we 
\begin{conj}
The bimodule $\Omega^1_D(C^\infty(\T^2)\rtimes \R)$ is generated
by $du_1$ and $du_2$ and is described by two relations,
\[
v_tdu_1=e^{iat}du_1v_t,~~v_tdu_2=e^{ibt}du_2v_t.
\]
\end{conj}
It seems that these difficulties in the end come from the quadratic part
in the signature operator $Q$. 

Let us note that we could choose another
diffeomorphism group, restricting the action of $\R$ to the subgroup $\Z$.
Then, the generators $V_t$ (or $v_t$) would be reduced to one generator
$V_1=V$ ($v_1=v$), and all the above formulae remain, replacing always $V_t$ ($v_t$)
by some power of $V$ ($v$).
However, we would not get rid of the difficulties related to the
differential calculus.

\section*{Acknowledgements}
The authors are grateful to K. Schm{\"u}dgen for helpful discussions.
R. M. was supported by the Deutsche Forschungsgemeinschaft. Part of
the work was done during his stay at the Max-Planck-Institute for
Mathematics in the Sciences in Leipzig.

\begin{appendix}

\section{Proof of Lemma \ref{lem3}}\label{appa}

\renewcommand{\theequation}{\Alph{section}.\arabic{equation}}
\setcounter{equation}{0}

The proof of this lemma rests on the following characterization of 
functions $f$ defined by determinants of Hankelian type, 
see~\cite{veindale1}, such that 
\begin{equation}\label{3.85}
\left|
\begin{array}{ccc}
f(i_0) & \cdots & f(i_0+k)\\
\vdots & \ddots & \vdots\\
f(i_{k}) & \cdots & f(i_{k}+k)
\end{array}
\right|= 0\enspace,
\end{equation}
$\forall k\in\N$ and $i_0,\dots,i_{k}\in\Z$. We have
\bth
A function $f$ defined on $\Z$ fulfils (\ref{3.85}) if and only if it
is of one of the following two types
\begin{eqnarray}
f_1(i)&=&\beta^i\sum\limits_{j=0}^{k-1}\alpha_j\,i^j\label{ap02}\\
f_2(i)&=&\sum\limits_{j=1}^k\alpha_j\beta_j^i\label{ap01}\enspace,
\end{eqnarray}
with $\alpha$, $\beta$ and $\beta_j\in\C$.
\ethe
\bpf
Let us first show by induction that $f_1$ and $f_2$ fulfil (\ref{3.85}). 
For $f_1(i)=\beta^i\sum\limits_{j=0}^{k-1}\alpha_j\,i^j$ and
$k=1$ we have
\[
\left|
\begin{array}{cc}
\alpha_0\beta^i & \alpha_0\beta^{i+1}\\
\alpha_0\beta^{j} & \alpha_0\beta^{j+1}
\end{array}
\right|= 0\enspace,
\]
$\forall i,j\in\Z$. Let, now (\ref{3.85}) be valid for $k=n$. Then we have
for $k=n+1$
\begin{eqnarray*}
&&\left|
\begin{array}{cccc}
\beta^{i_0}\sum\limits_{j=0}^{n}\alpha_ji_0^j &
\beta^{i_0+1}\sum\limits_{j=0}^{n}\alpha_j(i_0+1)^j& \cdots &
\beta^{i_0+n+1}\sum\limits_{j=0}^{n}\alpha_j(i_0+n+1)^j\\
\beta^{i_1}\sum\limits_{j=0}^{n}\alpha_ji_1^j &
\beta^{i_1+1}\sum\limits_{j=0}^{n}\alpha_j(i_1+1)^j& \cdots &
\beta^{i_1+n+1}\sum\limits_{j=0}^{n}\alpha_j(i_1+n+1)^j\\
\vdots & \vdots & \ddots & \vdots \\
\beta^{i_{n+1}}\sum\limits_{j=0}^{n}\alpha_ji_{n+1}^j &
\beta^{i_{n+1}+1}\sum\limits_{j=0}^{n}\alpha_j(i_{n+1}+1)^j& \cdots &
\beta^{i_{n+1}+n+1}\sum\limits_{j=0}^{n}\alpha_j(i_{n+1}+n+1)^j
\end{array}
\right|=\\
&&
\beta^{i_0+\cdots+i_{n+1}+\frac{(n+1)(n+2)}{2}}
\left|
\begin{array}{cccc}
\sum\limits_{j=0}^{n}\alpha_ji_0^j & 
\sum\limits_{j=0}^{n}\alpha_j(i_0+1)^j& \cdots &
\sum\limits_{j=0}^{n}\alpha_j(i_0+n+1)^j\\
\sum\limits_{j=0}^{n}\alpha_ji_1^j &
\sum\limits_{j=0}^{n}\alpha_j(i_1+1)^j& \cdots &
\sum\limits_{j=0}^{n}\alpha_j(i_1+n+1)^j\\
\vdots & \vdots & \ddots & \vdots \\
\sum\limits_{j=0}^{n}\alpha_ji_{n+1}^j &
\sum\limits_{j=0}^{n}\alpha_j(i_{n+1}+1)^j& \cdots &
\sum\limits_{j=0}^{n}\alpha_j(i_{n+1}+n+1)^j
\end{array}
\right|\enspace.
\end{eqnarray*}
But
\begin{eqnarray*}
&&\left|
\begin{array}{cccc}
\sum\limits_{j=0}^{n}\alpha_ji_0^j & 
\sum\limits_{j=0}^{n}\alpha_j(i_0+1)^j& \cdots &
\sum\limits_{j=0}^{n}\alpha_j(i_0+n+1)^j\\
\sum\limits_{j=0}^{n}\alpha_ji_1^j &
\sum\limits_{j=0}^{n}\alpha_j(i_1+1)^j& \cdots &
\sum\limits_{j=0}^{n}\alpha_j(i_1+n+1)^j\\
\vdots & \vdots & \ddots & \vdots \\
\sum\limits_{j=0}^{n}\alpha_ji_{n+1}^j &
\sum\limits_{j=0}^{n}\alpha_j(i_{n+1}+1)^j& \cdots &
\sum\limits_{j=0}^{n}\alpha_j(i_{n+1}+n+1)^j
\end{array}
\right|=\\
&&\left|
\begin{array}{cccc}
\sum\limits_{j=0}^{n}\alpha_ji_0^j & 
\sum\limits_{j=1}^{n}\alpha_j\left[(i_0+1)^j-i_0^j\right]& \cdots &
\sum\limits_{j=1}^{n}\alpha_j\left[(i_0+n+1)^j-(i_0+n)^j\right]\\
\sum\limits_{j=0}^{n}\alpha_ji_1^j & 
\sum\limits_{j=1}^{n}\alpha_j\left[(i_1+1)^j-i_1^j\right]& \cdots &
\sum\limits_{j=1}^{n}\alpha_j\left[(i_1+n+1)^j-(i_1+n)^j\right]\\
\vdots & \vdots & \ddots & \vdots \\
\sum\limits_{j=0}^{n}\alpha_ji_{n+1}^j &
\sum\limits_{j=1}^{n}\alpha_j\left[(i_{n+1}+1)^j-i_{n+1}^j\right]& \cdots &
\sum\limits_{j=1}^{n}\alpha_j\left[(i_{n+1}+n+1)^j-(i_{n+1}+n)^j\right]
\end{array}
\right|=\\
&&\left|
\begin{array}{cccc}
\sum\limits_{j=0}^{n}\alpha_ji_0^j & \sum\limits_{j=1}^n\alpha_j
\sum\limits_{l=0}^{j-1}
\left(j\atop l\right)i_0^l & \cdots &
\sum\limits_{j=1}^n\alpha_j\sum\limits_{l=0}^{j-1}
\left(j\atop l\right)(i_0+n)^l\\
\sum\limits_{j=0}^{n}\alpha_ji_1^j & \sum\limits_{j=1}^n\alpha_j
\sum\limits_{l=0}^{j-1}
\left(j\atop l\right)i_1^l & \cdots &
\sum\limits_{j=1}^n\alpha_j\sum\limits_{l=0}^{j-1}
\left(j\atop l\right)(i_1+n)^l\\
\vdots & \vdots & \ddots & \vdots\\
\sum\limits_{j=0}^{n}\alpha_ji_{n+1}^j & \sum\limits_{j=1}^n
\alpha_j\sum\limits_{l=0}^{j-1}
\left(j\atop l\right)i_{n+1}^l & \cdots &
\sum\limits_{j=1}^n\alpha_j\sum\limits_{l=0}^{j-1}
\left(j\atop l\right)(i_{n+1}+n)^l
\end{array}
\right|=\\
&&\sum\limits_{j=0}^{n}\alpha_ji_0^j
\left|
\begin{array}{ccc}
\sum\limits_{j=1}^n\alpha_j\sum\limits_{l=0}^{j-1}
\left(j\atop l\right)i_1^l & \cdots &
\sum\limits_{j=1}^n\alpha_j\sum\limits_{l=0}^{j-1}
\left(j\atop l\right)(i_1+n)^l\\
\vdots & \ddots & \vdots\\
\sum\limits_{j=1}^n\alpha_j\sum\limits_{l=0}^{j-1}
\left(j\atop l\right)i_{n+1}^l & \cdots &
\sum\limits_{j=1}^n\alpha_j\sum\limits_{l=0}^{j-1}
\left(j\atop l\right)(i_{n+1}+n)^l
\end{array}
\right|+\cdots+\\
&&(-1)^{n+1}\sum\limits_{j=0}^{n}\alpha_ji_{n+1}^j
\left|
\begin{array}{ccc}
\sum\limits_{j=1}^n\alpha_j\sum\limits_{l=0}^{j-1}
\left(j\atop l\right)i_0^l & \cdots &
\sum\limits_{j=1}^n\alpha_j\sum\limits_{l=0}^{j-1}
\left(j\atop l\right)(i_0+n)^l\\
\vdots & \ddots & \vdots\\
\sum\limits_{j=1}^n\alpha_j\sum\limits_{l=0}^{j-1}
\left(j\atop l\right)i_{n}^l & \cdots &
\sum\limits_{j=1}^n\alpha_j\sum\limits_{l=0}^{j-1}
\left(j\atop l\right)(i_{n}+n)^l
\end{array}
\right|=0\enspace,
\end{eqnarray*}
by assumption. Analogously, for 
$f_2(i)=\sum\limits_{j=1}^k\alpha_j\beta_j^i$ we have for $k=1$
\[
\left|
\begin{array}{cc}
\alpha_1\beta_1^i & \alpha_1\beta_1^{i+1}\\
\alpha_1\beta_1^j & \alpha_1\beta_1^{j+1}
\end{array}
\right|= 0\enspace,
\]
$\forall i,j\in\Z$. Let us now assume the validity of (\ref{3.85}) for $k=n$. 
Then we have for $k=n+1$
\begin{eqnarray*}
&&\left|
\begin{array}{cccc}
\sum\limits_{j=1}^{n+1}\alpha_j\beta_j^{i_0} &
\sum\limits_{j=1}^{n+1}\alpha_j\beta_j^{i_0+1} & \cdots &
\sum\limits_{j=1}^{n+1}\alpha_j\beta_j^{i_0+n+1}\\
\sum\limits_{j=1}^{n+1}\alpha_j\beta_j^{i_1} &
\sum\limits_{j=1}^{n+1}\alpha_j\beta_j^{i_1+1} & \cdots &
\sum\limits_{j=1}^{n+1}\alpha_j\beta_j^{i_1+n+1}\\
\vdots & \vdots &\ddots &\vdots\\
\sum\limits_{j=1}^{n+1}\alpha_j\beta_j^{i_{n+1}} &
\sum\limits_{j=1}^{n+1}\alpha_j\beta_j^{i_{n+1}+1} & \cdots &
\sum\limits_{j=1}^{n+1}\alpha_j\beta_j^{i_{n+1}+n+1}
\end{array}
\right|
=
\\
&&\left|
\begin{array}{cccc}
\sum\limits_{j=1}^{n+1}\alpha_j\beta_j^{i_0} &
\sum\limits_{j=1}^{n}\alpha_j\beta_j^{i_0}\left(\beta_j-
{\beta_{n+1}}\right) & \cdots &
\sum\limits_{j=1}^{n}\alpha_j\beta_j^{i_0+n}\left(
{\beta_j}-{\beta_{n+1}}\right)
\\
\sum\limits_{j=1}^{n+1}\alpha_j\beta_j^{i_1} &
\sum\limits_{j=1}^{n}\alpha_j\beta_j^{i_1}\left(\beta_j-
{\beta_{n+1}}\right) & \cdots &
\sum\limits_{j=1}^{n}\alpha_j\beta_j^{i_1+n}\left(
{\beta_j}-{\beta_{n+1}}\right)
\\
\vdots & \vdots &\ddots &\vdots\\
\sum\limits_{j=1}^{n+1}\alpha_j\beta_j^{i_{n+1}} &
\sum\limits_{j=1}^{n}\alpha_j\beta_j^{i_{n+1}}\left(
{\beta_j}-{\beta_{n+1}}\right) & \cdots &
\sum\limits_{j=1}^{n}\alpha_j\beta_j^{i_{n+1}+n}\left(
{\beta_j}-{\beta_{n+1}}\right)
\end{array}
\right|=\\
&&
\sum\limits_{j=1}^{n+1}\alpha_j\beta_j^{i_0}\left|
\begin{array}{ccc}
\sum\limits_{j=1}^{n}\alpha_j\beta_j^{i_1}\left(\beta_j^{}-
{\beta_{n+1}}\right) & \cdots &
\sum\limits_{j=1}^{n}\alpha_j\beta_j^{i_1+n}\left(
{\beta_j}-{\beta_{n+1}}\right)
\\
\vdots & \ddots & \vdots \\
\sum\limits_{j=1}^{n}\alpha_j\beta_j^{i_{n+1}}\left(
{\beta_j}-{\beta_{n+1}}\right) & \cdots &
\sum\limits_{j=1}^{n}\alpha_j\beta_j^{i_{n+1}+n}\left(
{\beta_j}-{\beta_{n+1}}\right)
\end{array}
\right|+\cdots+\\
&&(-1)^{n+1}
\sum\limits_{j=1}^{n+1}\alpha_j\beta_j^{i_{n+1}}\left|
\begin{array}{ccc}
\sum\limits_{j=1}^{n}\alpha_j\beta_j^{i_0}\left(\beta_j-
{\beta_{n+1}}\right) & \cdots &
\sum\limits_{j=1}^{n}\alpha_j\beta_j^{i_0+n}\left(
{\beta_j}-{\beta_{n+1}}\right)
\\
\vdots & \ddots & \vdots\\
\sum\limits_{j=1}^{n}\alpha_j\beta_j^{i_{n}}\left(
{\beta_j}-{\beta_{n+1}}\right) & \cdots &
\sum\limits_{j=1}^{n}\alpha_j\beta_j^{i_{n}+n}\left(
{\beta_j}-{\beta_{n+1}}\right)
\end{array}
\right|=0\enspace,
\end{eqnarray*}
by assumption.

Let us now assume that a function $f$ defined on $\Z$ fulfils 
(\ref{3.85}) for some $k\in\N$.
We choose $i_1=i_0+1,\dots,i_k=i_0+k$ and let $f(i_0),\dots,f(i_0+2k-1)$ 
denote the corresponding values of $f$. Then $f(i_0+2k)$ has to fulfil
\[
\left|
\begin{array}{cccc}
f(i_0) & f(i_0+1) & \cdots & f(i_0+k)\\
f(i_0+1) & f(i_0+2) & \cdots & f(i_0+k+1)\\
\vdots & \vdots &\ddots &\vdots\\
f(i_0+k) & f(i_0+1) & \cdots & f(i_0+2k)
\end{array}
\right|=0\enspace,
\]
provided that
\begin{equation}\label{3.89}
\left|
\begin{array}{cccc}
f(i_0) & f(i_0+1) & \cdots & f(i_0+k-1)\\
f(i_0+1) & f(i_0+2) & \cdots & f(i_0+k)\\
\vdots & \vdots &\ddots &\vdots\\
f(i_0+k-1) & f(i_0+k) & \cdots & f(i_0+2k-2)
\end{array}
\right|\not=0\enspace.
\end{equation}
(We may assume without loss of generality that (\ref{3.89}) holds. In 
\cite{ri1} it is shown that in the other case one is led to the case $k-1$.)
Proceeding further we find that the $2k$ values $f(i_0),\dots,f(i_0+2k-1)$ 
determine $f$ completely. Now we show that this function is
either of type (\ref{ap01}) or (\ref{ap02}).

Let first the constants $f(i_0),\dots,f(i_0+2k-1)$ be such that the 
following condition holds
\begin{equation}\label{ap3}
\sum\limits_{j=0}^{l+1}\beta^{l+1-j}f(i+j)(-1)^j\left(l+1\atop j\right)=0\enspace,
\end{equation}
for some $\beta\in\C$, $l\in\{0,\dots,k-1\}$ and all $i=i_0,\dots,i_0+2k-l-1$.
We show that the
corresponding function on $\Z$ is of the form (\ref{ap02}).
Suppose that $\beta\in\C$ is a solution of (\ref{ap3}). 
Then we find constants $\alpha_i$ as follows. Defining 
$g(i):=\frac{f(i)}{\beta^i}$ ($\beta\not=0$), 
we can always find $\alpha_i$ ($i=0,\dots,l$) as solutions of the 
following linear system of equations
\begin{eqnarray*}
g(i_0)&=&\alpha_0+\alpha_1i_0+\cdots+\alpha_li_0^l\\
g(i_0+1)&=&\alpha_0+\alpha_1(i_0+1)+\cdots+\alpha_l(i_0+1)^l\\
&\vdots&\\
g(i_0+l)&=&\alpha_0+\alpha_1(i_0+l)+\cdots+\alpha_l(i_0+l)^l
\end{eqnarray*}
by
\[
\alpha_i=\frac{1}{\Delta}
\left|
\begin{array}{cccccccc}
1 & i_0 & \cdots & i_0^{i-1} & g(i_0)& i_0^{i+1} & \cdots & i_0^l\\
1 & i_0+1 & \cdots & (i_0+1)^{i-1} & g(i_0+1)& (i_0+1)^{i+1} & \cdots &
(i_0+1)^l\\
\vdots & \vdots &&&&&&\vdots\\
1& i_0+l &  \cdots&(i_0+l)^{i-1}& g(i_0+l)&(i_0+l)^{i+1}&\cdots&(i_0+l)^l
\end{array}
\right|
\]
with
\[
\Delta=
\left|
\begin{array}{cccc}
1 & i_0 & \cdots & i_0^l\\
1 & i_0+1& \cdots & (i_0+1)^l\\
\vdots&&\vdots &\vdots \\
1&i_0+l&\cdots&(i_0+l)^l
\end{array}
\right|
=(-1)^{\frac{l(l+1)}{2}}\prod\limits_{j=1}^lj!\not=0\enspace.
\]
Now that we have chosen the constants $\beta$ and $\alpha_0,\dots,\alpha_l$ such
that
\[
f(i_0+j)=f_1(i_0+j)
\]
is fulfilled, for all $j=0,\dots,l$, it remains to be shown that we also have
\[
f(i_0+l+1)=f_1(i_0+l+1)\enspace.
\]
But
\[
\sum\limits_{j=0}^r(-1)^j\left(r\atop j\right)j^s=0\enspace,
\]
$\forall s=0,\dots,r-1$ (which follows from evaluating the $s$-th 
derivative of $f(x)=(x-1)^r=\sum\limits_{j=0}^r\left(r\atop j\right)(-1)^j
x^{r-j}$ at $x=1$). Now we find
\begin{eqnarray*}
\sum\limits_{j=0}^{l+1}\beta^{l+1-j}(-1)^j\left(l+1\atop j\right)f_1(i+j)&=&
\sum\limits_{j=0}^{l+1}\beta^{l+1-j}\beta^{i+j}(-1)^j\left(l+1\atop
j\right)\sum\limits_{m=0}^l\alpha_m(i+j)^m\\
&=&\beta^{l+i+1}\sum\limits_{j=0}^{l+1}\sum\limits_{m=0}^l\sum
\limits_{n=0}^m\alpha_m(-1)^j\left(l+1\atop j\right)
\left(m\atop n\right)i^nj^{m-n}\\
&=&\beta^{l+i+1}\sum\limits_{m=0}^l\alpha_m\sum\limits_{n=0}^m
\left(m\atop n\right)i^n\sum\limits_{j=0}^{l+1}(-1)^j\left(l+1\atop
j\right) j^{m-n}=0\enspace.
\end{eqnarray*}
Therefore, we have
\[
\sum\limits_{j=0}^{l+1}(-1)^j\beta^{l+1-j}\left(l+1\atop j\right)
\left(f(i_0+j)-f_1(i_0+j)\right)
=(-1)^{l+1}\left(f(i_0+l+1)-f_1(i_0+l+1)\right)=0\enspace,
\]
i.e.
\[
f(i_0+l+1)=f_1(i_0+l+1)\enspace.
\]
Let us now consider the general case (\ref{ap01}). Suppose, that 
$f(i_0),\dots,f(i_0+2k-1)$ are chosen such that (\ref{ap3}) does not hold. 
Then we have to solve the following system of algebraic equations
(where we have chosen $i_0=0$)
\begin{eqnarray*}
f(0)&=&C_1+\cdots+C_k\\
f(1)&=&C_1\beta_1+\cdots+C_k\beta_k\\
\vdots&&\\
f(2k-1)&=&C_1\beta_1^{2k-1}+\cdots+C_k\beta_k^{2k-1}\enspace.
\end{eqnarray*}
which can always be done using Gr{\"o}bner basis techniques, see~\cite{ri1}.
\epf\\[1ex]
\bre
If the parameters $f(i_0),\dots,f(i_0+2k-1)$ satisfy
\[
f(i_0+l)=\frac{f(i_0+1)^l}{f(i_0)^{l-1}}\enspace,
\]
$\forall l=0,\dots,2k-1$, then one easily checks that
\[
\beta=\frac{f(i_0+1)}{f(i_0)}
\]
fulfills (\ref{ap3}) and the constants $\alpha_i$ are given by
\[
\alpha_0=\frac{f(i_0)^{i_0+1}}{f(i_0+1)^{i_0}}\enspace,
\enspace\enspace\alpha_1=\cdots=\alpha_{k-1}=0\enspace.
\]
\ere
The proof of Lemma \ref{lem3} follows now immediately from the observation
that the function 
\[
h(i)=\sqrt{\lambda_{i0}}\gamma_{i0}-\sqrt{\lambda_{i-1,0}}\gamma_{i-1,0}\enspace,
\]
is obviously not of the form (\ref{ap02}) or (\ref{ap01}). 
\epf

\section{The differential algebra for the irrational rotation algebra}
\setcounter{equation}{0}

Let us first recall, see~\cite{rieffel1,co8.5}, that the algebra of the 
noncommutative torus is generated by two unitaries $u, v$ subject to 
the relation
\[
uv={\rm e}^{-2\pi {\rm i}\theta}vu\enspace.
\]
The algebra can be considered on the purely *-algebraic level (Laurent
polynomials in $u,~v$) where a general element is a finite linear
combination of ordered polynomials $u^kv^l$, $k,l\in\Z$, on the level
of smooth functions, where the general element is a series
$\sum a_{kl}u^kv^l$ with coefficients $a_{kl}$ subject to the condition
that $(|k|^n+|l|^n)|a_{kl}|$ are bounded for all $n>0$.
Finally, there is also the $C^*$-version, defined e.g. by using irreducible
representations for performing a norm closure of the polynomial algebra. 
It is well-known that both the polynomial and the $C^*$-algebra can
be interpreted as convolution algebras of the reduced holonomy groupoid
of the Kronecker foliation, whith $\theta$ being the angle defining the
direction of the leaves. We denote the $C^*$-algebra by $A_\theta$, the
smooth algebra by $\cA_\theta$ and the poynomial algebra by $\cO_\theta$.
There exists a tracial state $\tau$ on $A_\theta$, given by
\[
\tau(\sum a_{kl}u^kv^l)=a_{00},
\]
and there are two canonical derivations $\delta_1$ and $\delta_2$ on
$\cA_\theta$ defined by
\[
\delta_1(u^kv^l)=2\pi i ku^kv^l,~
\delta_2(u^kv^l)=2\pi ilu^kv^l.
\]
With these data, the well-known spectral triple is defined as follows:
First, the tracial state $\tau$ is used to define the GNS Hilbert space
$\cH_\tau$. Secondly, the derivations $\delta_1$ and $\delta_2$ give rise
to unbounded operators on $\cH_\tau$, whose domain of definition is the image
of $\cA_\theta$ in $\cH_\tau$ (under the GNS procedure). The same is true for
$\partial:= \frac{1}{\sqrt{2\pi}}(\delta_1-i\delta_2)$.
Now take $\cH:=\cH_\tau\oplus \cH_\tau$ and 
$D:=\left(\ba{cc}0&\partial\\\partial^*&0\ea\right)$
as Hilbert space and generalized Dirac operator of a spectral triple over
$\cA_\theta$. The dimension of this spectral triple is known to be two.
The corresponding differential calculus $\Omega_D$ was described by Connes
in terms of elements of $\cH$. We have the following description of
$\Omega_D(\cO_\theta)$ in terms of relations
between the generators
of the algebra and their differentials:
\bpr
(i) $\Omega^1_D(\cO_\theta)$ is a free left (or right) 
$\cO_\theta$-module with
basis $\{du,dv\}$. The bimodule structure of $\Omega^1_D(\cO_\theta)$
is given by
\beq\label{U}
udu=duu,~u^*du=duu^*,~udu^*=du^*u,~u^*du^*=du^*u^*,
\eeq
\beq\label{V}
vdv=dvv,~v^*dv=dvv^*,~vdv^*=dv^*v,~v^*dv^*=dv^*v^*,
\eeq
\beq\label{UV}
vdu=e^{2\pi i\theta}duv,~~udv=e^{-2\pi i\theta}duv,
\eeq
\beq\label{U*V}
vdu^*=e^{-2\pi i\theta}du^*v,~~u^*dv=e^{2\pi i\theta}dvu^*,
\eeq
\beq\label{V*U}
v^*du=e^{-2\pi i\theta}duv^*,~~udv^*=e^{2\pi i\theta}dv^*u.
\eeq
(ii) $\Omega^2_D(\cA_\theta)$ is a free left (or right) $\cA_\theta$-module
with basis $\{dudv\}$. The relation
\beq\label{dudv}
dudv=-e^{2\pi i\theta}dvdu
\eeq
is fulfilled.\\
(iii) $\Omega^k_D(\cA_\theta)=0$ for $k\geq 3$.
\epr
\bpf
(i) $\tau$ is a faithful state, thus the GNS representation $\pi$ is faithful.
Consequently, $\Omega^1_D(\cO_\theta)\simeq \pi(\Omega^1(\cO_\theta))$,
where the isomorphism sends differentials to commutators with $D$. To verify
the relations (\ref{U})--(\ref{V*U}) it is therefore sufficient to consider
the images of these expressions under $\pi$. If we denote by $\underline{a}$
the element corresponding to $a\in\cO_\theta$ in $\cH_\tau$, it is immediately 
verified that the $e_{kl}:=\underline{u^kv^l}$ form an orthonormal basis
in $\cH_\tau$. From this basis we obtain in an obvious way an orthonormal
basis $\{e^+_{kl},e^-_{kl}\}$ of $\cH_\tau\oplus \cH_\tau$ (to be precise,
$e^+_{kl}=(e_{kl},0),~e^-_{kl}=(0,e_{kl})$). In this basis,
$U:=\pi\oplus\pi(u)$,
$V:=\pi\oplus\pi(v)$, $D$ act as follows:
\beq\label{uv}
U(e^\pm_{kl})=e^\pm_{k+1,l},~~V(e^\pm_{kl})=e^{2\pi ik\theta}e^\pm_{k,l+1},
\eeq
\beq\label{D}
D(e^\pm_{kl})=\sqrt{2\pi}(\pm ik+l)e^\mp_{kl}.
\eeq
Now, it is straightforward to verify the relations (\ref{U})--(\ref{V*U})
(with $U, V, [D,\cdot]$ instead of $u,v,d$).
From the these and the Leibniz rule (also taking into
account unitarity of the generators $u,v$) it is obvious that 
$[D,U]$ and $[D,V]$ generate $\pi(\Omega^1(\cO_\theta))$ as a left (or
right) $\cO_\theta$-module.
To prove that it is a freely generated left module, assume $P[D,U]+Q[D,V]=0$ 
with $P,Q\in
\pi\oplus\pi(\cO_\theta)$. It follows from (\ref{uv}) and (\ref{D}) that
\begin{eqnarray}
\left[D,U\right](e^\pm_{kl}) &=& \pm i\sqrt{2\pi}e^\mp_{k+1,l}\\
\left[D,V\right](e^\pm_{kl}) &=
& e^{2\pi ik\theta}\sqrt{2\pi}e^\mp_{k,l+1}\enspace.
\end{eqnarray}
Therefore, terms in $P[D,U]+Q[D,V]$ can only compensate if they
contain the same overall number of $U$ and of $V$. This means that it is
sufficient to consider terms of the form
$\alpha=pU^nV^{m+1}[D,U] + qU^{n+1}V^m[D,V]$, $p,q\in\C$.
Acting on $e^+_{kl}$, we obtain
\[
\alpha e^+_{kl}=\sqrt{2\pi}(e^{2\pi i(k+1)(m+1)\theta}ip+e^{2\pi
ikm\theta}q)e^-_{k+n+1,l+m+1}=0,
\]
which is equivalent to
\[
pe^{2\pi i(k+m+1)}+q=0.
\]
Since this should be true for all $k$, it follows that $p=q=0$.\\
(ii) Differentiating the relations (\ref{U}) and (\ref{V}) gives immediately
$dudu=dvdv=du^*du^*=dv^*dv^*=0$. Analogously, (\ref{UV}) leads to
(\ref{dudv}). Thus, we know already that $dudv$ generates the two forms
as a left (or right) $\cA_\theta$-module. It remains to show that it is freely
generated.

Let us recall that
\[
\Omega^2_D(\cO_\theta)\simeq \pi(\Omega^2(\cO_\theta))/\pi(dJ^1),
\]
where $J^1=\ker \pi\cap\Omega^1(\cO_\theta)$. Thus, any relation true
in $\pi(\Omega^2(\cO_\theta))$ is also true in
$\pi(\Omega^2_D(\cO_\theta))$. 
Now, a similar argument as in the proof of Proposition \ref{oql}
shows that
(i) implies that $J^1$ coincides with the $\cO_\theta$-subbimodule generated
the elements corresponding to the relations (\ref{U})--(\ref{V*U}).
It follows that $dJ^1$ is a finite
sum of elements of the form $a~db~c$ with $a,c\in \cO_\theta$ and $b$ one of
the elements (\ref{U})--(\ref{V*U}). Now, one shows by a direct computation
that $\pi(db)\in\pi(\cO_\theta)$ if $b$ is one of the elements
(\ref{U})--(\ref{V}) whereas $\pi(db)=0$ if $b$ is one of the remaining
elements. It follows that $\pi(dJ^1)=\pi(\cO_\theta)$. It remains to show
that from $\pi(a)[D,U][D,V]\in\pi(\cO_\theta)$ it follows that $a=0$. From
the above formulae it is now immediate that any element of the algebra
acts on the $e^\pm_{kl}$ in a way not depending on $+$ or $-$,
$\pi(a)e^\pm_{kl}=\sum\lambda_{ij}e^\pm_{ij}$, $\lambda_{ij}$ independent
on + or -. On the other hand,
\[
[D,U][D,V]e^\pm_{kl}=\pm 2\pi i e^{2\pi ik\theta}e^\pm_{k+1,l+1}\enspace,
\]
from which (ii) follows immediately.\hfill$\square$
\bre
As in Remark \ref{topcal}, we can construct a topological version
$\Omega_D(\cA_\theta)$ of this calculus (using seminorms $q_n(\sum a_{kl}
u^kv^l)=\sup_{kl}(1+|k|^n+|l|^n)|a_{kl}|$). A comparison with the results
of \cite{s-a01} shows that this gives indeed the calculus
$\Omega_D(\cA_\theta)$ of the spectral triple $(\cA_\theta,\cH,D)$.
\ere
\end{appendix}

\end{document}